\documentclass[12pt]{amsart}

\usepackage{ljm-auth}

\usepackage{amssymb}
\usepackage{amsmath}
\usepackage{amsthm}
\usepackage{textcomp}
\usepackage{array}
\usepackage[utf8]{inputenc}
\usepackage[english]{babel}

\usepackage{graphicx}

\usepackage{hyperref}

\author{K.~Khadiev}
\crauthor{Khadiev} 
\tit{On the Hierarchies for Deterministic, Nondeterministic and Probabilistic Ordered  Read-$k$-times  Branching Programs}
\shorttit{On the Hierarchies for $k$-OBDD, $k$-NOBDD and $k$-POBDD} 

\newtheorem{theorem}{Theorem}
\newtheorem{lemma}{Lemma}
\newtheorem{definition}{Definition}
\newtheorem{remark}{Remark}
\newtheorem{corollary}{Corollary}
\newtheorem{property}{Property}
\newcommand{\Endproof}{\hfill$\Box$\\}
\newcommand{\Beginproof}{{\em Proof.\quad}}

\begin{document}

\maketit
\address{ Kazan Federal University,
Institute of Computational Mathematics and Information Technologies, Department of Theoretical Cybernetic,
     Kazan, Russia\\
     University of Latvia, Department of Computing, Riga, Latvia}

\email{kamilhadi@gmail.com}

\abstract{The paper examines hierarchies for nondeterministic and
deterministic ordered read-$k$-times Branching programs. The currently known hierarchies for deterministic $k$-OBDD models of Branching programs for $ k=o(n^{1/2}/\log^{3/2}n)$ are proved by  B.~Bollig, M.~Sauerhoff, D.~Sieling, and I.~Wegener  in 1998. Their lower bound technique was based on communication complexity approach. For nondeterministic $k$-OBDD it is known that, if $k$ is constant then polynomial size $k$-OBDD computes same functions as  polynomial size OBDD (The result of Brosenne, Homeister and Waack, 2006). In the same time currently known hierarchies for nondeterministic read $k$-times Branching programs for $k=o(\sqrt{\log{n}}/\log\log{n})$ are proved by Okolnishnikova  in 1997, and  for probabilistic read $k$-times Branching programs for $k\leq \log n/3$ are proved by Hromkovic and Saurhoff  in 2003.        

We show that increasing $k$ for polynomial size nodeterministic $k$-OBDD makes model more powerful if $k$ is not constant. Moreover, we extend the hierarchy  for probabilistic and nondeterministic $k$-OBDDs for $ k=o(n/ \log n)$. These results extends hierarchies for read $k$-times Branching programs, but $k$-OBDD has more regular structure. The lower bound techniques we propose are a ``functional description'' of Boolean function presented by nondeterministic $k$-OBDD and communication complexity technique. We present similar hierarchies for  superpolynomial and subexponential width nondeterministic $k$-OBDDs.

Additionally  we expand the  hierarchies for deterministic $k$-OBDDs using our lower bounds  for $ k=o(n/ \log n)$.   We also analyze similar hierarchies for superpolynomial and subexponential width $k$-OBDDs.}

\notes{0}{
}

\section{Preliminaries and Results}
Ordered  Read $k$-times  Branching Programs ($k$-OBDD) are well known models for Boolean functions computation. A good source for different models of branching programs is the book by Ingo Wegener  \cite{Weg00}.

A branching program  over a set $X$ of $n$ Boolean variables is
a directed acyclic graph with two distinguished nodes $s$ (a source node) and $t$ (a sink node). We denote  such a program as $P_{s,t}$ or just $P$.   Each inner node $v$  of $P$ is associated with a variable $x\in X$. {\em Deterministic} $P$ has exactly two outgoing edges labeled $x=0$   and $x=1$ respectively; for a such node $v$,  {\em nondeterministic}   $P$  has several  outgoing edges labeled $x=0$   and $x=1$ respectively.

The program $P$ (deterministic or nondeterministic) computes the Boolean function $f(X)$ ($f:\{0,1\}^n \rightarrow \{0,1\}$) as follows: for each $\sigma\in\{0,1\}^n$ we let $f(\sigma)=1$ if and only if there exists at least one $s-t$ path (called {\em accepting} path for $\sigma$) such that all edges along this path are consistent with $\sigma$.

Branching program $P$ is called {\em syntactic read-$k$-times}
BP if for any path (consistent  or inconsistent)  from the source to a
sink node of $P$ the program reads each variable at most $k$ times \cite{brs93}.

A branching program is {\em leveled} if the nodes can be partitioned into levels $V_1, \ldots, V_\ell$ and a level $V_{\ell+1}$ so that the nodes in $V_{\ell+1}$ are the sink nodes and nodes in each level $V_j$ with $j \le \ell$ have outgoing edges only to nodes in the next level $V_{j+1}$. For a leveled $P_{s,t}$ the source node $s$ is a node from the first level  $V_1$ whereas the sink node $t$ is a node from the last level $V_\ell$ of nodes.

The {\em width} $w(P)$ of a leveled branching program $P$ is the maximum
of number of nodes in  levels of $P$
\[ w(P)=\max_{1\le j\le \ell}|V_j|. \]

A leveled branching program is called {\em oblivious} if all inner
nodes of one level are labeled by the same variable.  A branching
program is called {\em read once} if each variable is tested on each
path only once.

An oblivious leveled read once branching program is also called Ordered  Binary Decision Diagram (OBDD) and for nondeterministic case is NOBDD.

OBDD (NOBDD) $P$ reads variables in its individual  order
$\theta(P)=(j_1,\dots,j_n)$. We call $\theta(P)$ the order of $P$.

The Branching program $P$ is called $k$-OBDD ($k$-NOBDD)  if it  consists on $k$ layers, where  $i$-th ($1\le i\le k$) layer  $P^i$ of $P$  is  an OBDD (NOBDD). Let  $\theta_i$ be an order of $P^i$, $1\le i\le k$.
And $\theta_1=\dots=\theta_k=\theta$.
 We call  order
$\theta(P)=\theta$ the order of $P$. Therefore we conclude that $k$-OBDD is a  specific case of   syntactic read-$k$-times
BP. 

The {\em size} $S(P)$ of branching program $P$ is a number of  nodes of
program $P$. Note, that for $k$-OBDD ($k$-NOBDD) we have
$ S(P) \le w(P) \cdot n \cdot k $

The {\em length} $l(P)$ of branching program $P$ is the length of the longest path from source to a sink node. Clearly we have, that for $k$-OBDD ($k$-NOBDD) $P$ its length $l(P)$ is  $n \cdot k$

We can consider probabilistic $k$-OBDD by the same way. Vertexes of  this model's graph can have more that two outgoing edges and we choose the edge according to probabilistic mechanism. We grantee that computation will be finished with probability $1$.

In the paper will be considered bounded error model. Program returns $1$ on input $\nu\{0,1\}^n$ with bounded error $\delta$ if probability $\Pr\{R\mbox{ reaches }1\mbox{ on }\nu\}>0.5+\delta$. In that case $P(\nu)=1$. And returns $0$ if  $\Pr\{R\mbox{ reaches }1\mbox{ on }\nu\}<0.5-\delta$. In that case $P(\nu)=0$.

{\bf Known lower bounds and  hierarchies.}
Let P-$k$BP be the set of Boolean functions that can
be computed by syntactic read-$k$-times BP of  polynomial size, and NP-$k$BP  be the set of Boolean functions for nondeterministic case.
Let P-$k$OBDD be the set of Boolean
functions that can  computed  by  $k$-OBDD of polynomial size, and NP-$k$OBDD be the set of Boolean functions for nondeterministic case.

One of the first explicit hard functions  for the {\em syntactic}
BPs was introduced in  \cite{brs93} by  Borodin, Razborov and
Smolensky. For each $k \leq c \log n$ they presented an explicit
function, which needs  non-polynomial size syntactic $k$-BPs for some
appropriate constant $c>0$.

 Thathachar in paper \cite{tha98} presented a family of explicit Boolean
functions depending on integer parameter $k$  which cannot be
represented as $kn$ length  polynomial size {\em syntactic} nondeterministic k-BP.
In addition, the technique from \cite{tha98} allows to prove the following proper inclusion  for $k=o(\log\log n)$:
 \[\mbox{NP-}(k-1)\mbox{BP} \subsetneq  \mbox{NP-}k\mbox{BP},\mbox{ for }k=o(\log\log n).\]

 This result extends the result of Okolnishnikova \cite{oko97} that proved the  following hierarchy:
 \[\mbox{NP-}k\mbox{BP}\subsetneq \mbox{NP-}(k\ln k/2 + C)\mbox{BP},\mbox{ for }k=o(\sqrt{\ln{n}}/\ln\ln{n}).\]
 
 Probabilistic $k$-BP was investigated by Hromkovich and Sauerhoff in 2003 \cite{hs2003}. They proved lower bound for explicit Boolean function $m$-$Masked$-$PJ_{k,n}$. Authors showed that bounded error probabilistic $k$-BP should have a size at least $2^{\Omega(N^\alpha/k^3)}$, for $\alpha = 1/(1 + 2 log 3)$. Using that results  Hromkovich and Sauerhoff got hierarchy for polynomial size bounded error  probabilistic $k$-BP: BPP-$(k-1)$BP $\subsetneq$ BPP-$k$BP, for $k\leq\log n/3$.

What is known for the  read-$k$-times BP models with an extra ``ordered reading'' restrictions?

For the case of  $k$-OBDD models Bolling,
Sauerhoff, Sieling, Wegener suggested an explicit Boolean function
which cannot be represented by  non-linear length  $o(n^{3/2}/\log^{3/2}n)$
 polynomial size $k$-OBDDs. In
addition their technique allows to prove the following
proper inclusions, for
$k=o(n^{1/2}/\log^{3/2}n)$
\[\mbox{P-}(k-1)\mbox{OBDD} \subsetneq \mbox{P-}k\mbox{OBDD}.\]

For small width $k$-OBDD we presented lower bounds in paper \cite{ak13}, which allows to extends the Bolling--Sauerhoff--Sieling--Wegener hierarchy for sublinear width, similar to width hierarchy which was proved in \cite{k2015}.

Ablayev and Karpinski in  \cite{ak96} introduced an explicit Boolean function which is hard for polynomial size nondeterministic $k$-OBDD for $k=o(n/\log n)$, but can be  computed by bounded-error  probabilistic  $k$-OBDD. In another word the function $f$ from  \cite{ak96} has the following property: $f\in \mbox{coRP-}k\mbox{OBDD} \backslash \mbox{NP-}k\mbox{OBDD}$.

Brosenne,  Homeister and Waack \cite{bhw2006} showed that for any constant $k$ holds $\mbox{NP-}\mbox{OBDD} = \mbox{NP-}k\mbox{OBDD}$.

{ \bf Our contribution.}
In Sections \ref{s-lower} and \ref{sec:hierarchy} we consider Boolean function $EQS_{d}$ ({\em Shuffled Equality}) which is presented in \cite{agky14}, this is modification of $EQS$ function from \cite{ak96}, \cite{a97}, \cite{A97ICALP}. We prove  that $EQS_{d}$  cannot be
represented by polynomial size $k$-NOBDDs for $k=o(n/\log n)$. Our technique is the so called ``functional description'' of Boolean function presented in a corresponding $k$-NOBDD. Namely,    we develop a technique of
presentation  $k$-NOBDD  as special decomposition, which presented in Section \ref{s-lower}. This technique is motivated by the paper \cite{brs93}. 


Based on our lower bound in Section \ref{sec:hierarchy} we prove the following proper
inclusion (Corollary \ref{hierarchyResult1} and \ref{hierarchyResult2}). For $k = o(n/\log n)$
and $k>\log^2 n$ it holds that
\[ \mbox{P-}\left(k/\log^2 n\right)\mbox{OBDD}\subsetneq \mbox{P-}k\mbox{OBDD}, \quad \mbox{NP-}\left(k/\log^2 n\right)\mbox{OBDD}\subsetneq \mbox{NP-}k\mbox{OBDD}. \]

These results presented by Theorems \ref{th-hi} and \ref{th-hi2} and for polynomial size $k$-OBDD. The hierarchy is based on $EQS_d$ function.   Corollary \ref{hierarchyResult1} for nondeterministic case and Corollary \ref{hierarchyResult2} for deterministic one.

 Additionally, we present the  hierarchies  for superpolynomial and subexponential sizes (Corollary \ref{hierarchyResult1} and \ref{hierarchyResult2}).

The result for deterministic case extends Bolling--Sauerhoff--Sieling--Wegener hierarchy in polynomial case. Hierarchies for superpolynomial and subexponential size are new. In nondeterministic case we show that increasing $k$ for nodeterministic $k$-OBDD makes model more powerful, if $k$ is not constant. And prove hierarchy which extends Thathachar's and Okolnishnikova's  hierarchies but for model with more regular structure ($k$-OBDD). 

Second group of results is based on result from communication complexity. Namely, we develop a technique of
presentation  $k$-OBDD  as special communication protocol. We take this result from \cite{ak13} and prove similar results for nondeterministic and probabilistic $k$-OBDDs in the paper. 
In Sections \ref{s-lower-small} and \ref{s-lower-p-small} we consider Boolean function $SAF_{k,w}(X)$ ({\em Shuffled Address Function}) which is presented in \cite{k2015}. We prove that $SAF_{k,w}(X)$  cannot be
represented by constant width $k$-OBDDs for $k=o(n/\log n)$ and sublinear width  $k$-OBDDs for $k=o(n^{1-\alpha}/\log n),0 < \alpha < 0.49$. 

Based on our lower bound in Section \ref{sec:hierarchy} we prove the following proper
inclusion (Corollary \ref{hierarchyResult1-small}, \ref{hierarchyResult2-small}) and \ref{hierarchyResult3-small}).


These results presented by Theorems \ref{th-hi-small}, \ref{th-hi2-small} and \ref{th-hi3-small}. The hierarchy is based on $SAF_{k,w}(X)$ function.   Corollary \ref{hierarchyResult1-small} for nondeterministic case, Corollary \ref{hierarchyResult2-small} for deterministic case and Corollary \ref{hierarchyResult3-small} for probabilistic case.

The result for deterministic case extends Bolling--Sauerhoff--Sieling--Wegener hierarchy for sublinear width. In nondeterministic case we prove hierarchy which extends Thathachar's and Okolnishnikova's  hierarchies but for model with more regular structure ($k$-OBDD) and sublinear width.  For probabilistic case we prove hierarchy which extends Hromkovich and Saurhoff  hierarchy but for model with more regular structure ($k$-OBDD) and sublinear width.

These two groups of results, which based on ``functional description'' and communication complexity technique, complement each other. First group of results cannot be applied for sublinear width, but second one cannot be applied for polynomial width. The both groups extend Bolling--Sauerhoff--Sieling--Wegener hierarchy in different cases.

\section{Decomposition and Simulation of Nondeterministic Ordered  Read-$k$-times  Branching Programs}\label {s-lower}

\subsection{Decomposition of Nondeterministic Ordered  Read-$k$-times  Branching Programs}

We denote by ${\bf NOBDD_w}$ the class of Boolean functions that are computable by NOBDDs of width $w$.

We denote by $ k{\bf\mbox{-}NOBDD}_w$ the class of Boolean functions that are computable by $k$-NOBDDs of width $w$.

Let ${\bf F}(d,q,r)$ be a set of Boolean functions, over $X=\{x_1,\dots,x_n\}$, which are presented in the following form:
\begin{equation}
\label{dmainkobddbound}
{\bf F}(d,k,w)=\left\{g(X)=\bigvee_{j= 1}^{d}\bigwedge_{i= 1}^{k} g_{j,i}(X)\right\},
\end{equation}
where $g_{j,i}\in {\bf NOBDD}_w$.

\begin{lemma}\label{th-main1} For integer $k,w$, such that $k\log w<n$, the following statement is true:
 $ k{\bf\mbox{-}NOBDD}_w\subseteq {\bf F}(w^{k-1}, k, w)$.
\end{lemma}
%
%
{\em Proof.}  Let $P^i$ be an  $i$-th layer ($1\leq i \leq k$) of $P_{s,t}$. $P^i$ is an $n$ leveled NOBDD. Let  $V^i_1$   be a set of nodes of the first level of  $P^i$  and $V^i_n$  be  a set of nodes of the $n$-th (last) level of  $P^i$. Clearly we have the situation when for $1\le i\le k-1$ the $n$-th level $V^i_n$ of nodes of $P^i$ is coincide with the first level $V^{i+1}_1$ of nodes of $P^{i+1}$, that is $V^{i}_n=V^{i+1}_1$.

We call a sequence $m_1,\dots,m_{k+1}$ of nodes of program $P_{s,t}$ a {\em trace} and denote it as $tr(m_1,\dots,m_{k+1})$ if and only if the following statements are true:
\begin{enumerate}\label{trace}
\item $m_1=s$,
\item $m_i\in V^i_1$ for $2\le i\le k$,
\item $m_{k+1}=t$.
\end{enumerate}
It is easy to see that any $s-t$ path $p$ of the program $P_{s,t}$ contains a (uniquely determined) trace $tr(m_1,\dots,m_{k+1})$ where nodes $m_1,\dots,m_{k+1}$ appear along $p$ in this prescribed order. We denote by $TR$ the set of all traces of program $P_{s,t}$.

For a trace $tr(m_1,\dots,m_{k+1})$, for $1\le i\le k$
by  $g_{m_i,m_{i+1}}(X)$ we denote the  function computed by the program $P^i_{m_i,m_{i+1}}$.

%
%

Now we define the following function $g(X)$:
\begin{equation}\label{dec-ibdd}
 g(X)=  \bigvee_{tr(m_1,\dots,m_{k+1})\in TR}  \bigwedge_{i=1}^{k} g_{m_i,m_{i+1}}(X).
 \end{equation}

This function expresses the fact that there is at least one trace $tr(m_1,\dots,m_{k+1})\in TR$ and at least one accepting path $p$ for the input $\sigma\in f^{-1}(1)$ being considered such that $p$ contains trace $tr(m_1,\dots,m_{k+1})$. Hence, $g=f$ and we only have to check that the representation (\ref{dec-ibdd}) has the desired form (\ref{dmainkobddbound}).

Indeed, the function $g_{m_i,m_{i+1}}(X)$ is computed by the program $P^i_{m_i,m_{i+1}}$ which is a NOBDD  of width $w$ with the source $m_i$ and sink $m_{i+1}$. Hence, $g_{m_i,m_{i+1}}(X)\in {\bf NOBDD}_w$.

Furthermore, according to the definition of trace, the total number of traces does not exceed $w^{k-1}$:
\[ |TR|\le w^{k-1}. \]
These two facts complete the proof of the lemma. \Endproof

\subsection{Constructing Nondeterministic Ordered  Read-ones  Branching Program by Decomposition of Nondeterministic Ordered  Read-$k$-times  Branching Program}

In this section we show that we can associate decomposition of Boolean function with computing by NOBDD.

\begin{theorem}\label{similate}
For integer $k,w,d$, such that $k\log w<n$, following statement is true:
\[  k{\bf\mbox{-}NOBDD}_w\subseteq{\bf NOBDD}_{w^{2k-1}} .\]
\end{theorem}
{\em Proof.}
 Informally speaking we prove that the $k$-NOBDD $P$ can be simulated by NOBDD $R$ of lager width than $P$.
Note that the idea of such simulation is folklore and is used, for example, in \cite{hskks02}, \cite{h97}.

Let us present it formally.

Let $g\in k{\bf\mbox{-}OBDD}_w$, then $P$ is computed by $k$-NOBDD $P$ of width $w$. According to Lemma \ref{th-main1} we have $g\in{\bf F}(w^k,k,w)$, hence the function is introduced in the following form:
\[
 g(X)=  \bigvee_{tr(m_1,\dots,m_{k+1})\in TR}  \bigwedge_{i=1}^{k} g_{m_i,m_{i+1}}(X)
\]
for $g_{m_i,m_{i+1}}\in {\bf NOBDD}_w$.

In order to show that $g\in {\bf NOBDD}_{w^{2k-1}}$ we construct NOBDD $R$ of with $w^{2k-1}$.

The graph of $R$ consists from $|TR|$ parallel parts, which are chosen in nondeterministic way in the first step. Each part is associated with different parts for different traces $T\in TR$ and we denote it $R(T)$.

Let  $P(T)$, for the trace $T=tr(m_1,\dots,m_{k+1})$,  is subprogram of $P$ determined by exactly all  $s-t$ paths that contains trace T. Note that $w(P(T))=w$.

First of all,  we show that we can simulate $P(T)$ for trace $T=tr(m_1,\dots,m_{k+1})$ with NOBDD $R(T)$ of width $w^k$, note that $P(T)$ computes boolean function $ \bigwedge_{i=1}^{k} g_{m_i,m_{i+1}}(X)$.

A graph of $R(T)$ contains $n+1$ levels of nodes $W_1,\dots,W_{n+1}$.
The first level of $R(T)$ is $W_1=V_{1}\times V_{n+1}\times \dots \times V_{(k-1)n+1}$=$\{(m_1,\dots,m_k)\}$, where $V_i$ is $i$-th level of $P(T)$.

The $i$-th level of $R(T)$ is $W_i=V_{i}\times V_{n+i}\times V_{2n+i} \times \dots \times V_{(k-1)n+i}$  for $2\leq i \leq n$.

Last level of $R(T)$ is $W_{n+1}=V_{n+1}\times V_{2n+1}\times \dots \times V_{kn+1}$=$\{(m_2,\dots,m_{k+1})\}$.

Transition from state $v\in W_i$ under input $0$ $(1)$ is determined in transitions of $P(T)$ under input $0$ $(1)$ on the levels $i,n+i,\dots (k-1)n+i$.
 
Obviously, that for some $\sigma\in\{0,1\}^n$ NOBDD $R(T)$ reaches terminate node if $P(T)$ reaches terminate node and it means that program $P$ has used trace $T$.

Also we have $w(R(T))=w^k$ by definition of $R(T)$.

It should be pointed out that $R$ contains all $R(T)$. In the first step $R$ chooses one of $T\in TR$ in nondetermionistic way and then work according to $R(T)$. Moreover, note that $|TR|=w^{k-1}$, hence $w(R)=w^{2k-1}$.

The NOBDD $R$ computes Boolean function $g$, due to (\ref{dec-ibdd}).
\Endproof

\subsection{Lower bound for Boolean Function $EQS_d$.}

Let set $C$ be one of the following sets:
\begin{itemize}

\item
${\tt poly}=\{w:$ $w$ is polynomial$, w>n^2\}$. It means that $k$-OBDD ($k$-NOBDD) $P$ has polynomial size.

\item
${\tt superpoly}_\alpha=\{w: w=O(n^{log^{\alpha}n})\}$, $\alpha>0$  and it means that $k$-OBDD ($k$-NOBDD) $P$ has super polynomial size.

\item
${\tt subexp}_\alpha=\{w: w=O(2^{n^{\alpha}})\}$, $0<\alpha<0.5$,  and it means that $k$-OBDD ($k$-NOBDD) $P$ has subexponential size.
\end{itemize}

Let  $  k{\bf\mbox{-}OBDD}_C$ and $  k{\bf\mbox{-}NOBDD}_C$ be the set of Boolean functions that have representation as $k$-OBDD and $k$-NOBDD with width $w\in C$, respectively.

We consider Boolean function $EQS_d$, which was defined in \cite{agky14}, as a modification of  Boolean function  {\em Shuffled Equality} which was defined in \cite{ak96} and \cite{a97}.

{ \bf Function $EQS_d$.}
Let $d$ be multiple of $4$ such that $4\leq d\leq 2^{n/4}$. The Boolean function $EQS_d$ depends only on the first $d$ bits.

For any given input $\nu\in\{0,1\}^n$, we define two binary strings $\alpha(\nu)$ and $\beta(\nu)$ in the following way. We call odd bits of the input {\em marker bits} and even bits  {\em value bits}. For any $i$ satisfying $1\leq i \leq d/2$, the value bit $\nu_{2i}$ belongs to $\alpha(\nu)$  if the corresponding marker bit $\nu_{2i-1}=0$ and $\nu_{2i}$ belongs to $\beta(\nu)$ otherwise.

\begin{displaymath}
EQS_d(\nu) = \left\{ \begin{array}{ll}
1, & \textrm{if } \alpha(\nu)= \beta(\nu);\\
0, & \textrm{otherwise.}
\end{array} \right. 
\end{displaymath}


\begin{lemma}\label{peq2_knobdd}
$EQS_k\not\in (k/r){\bf\mbox{-}NOBDD}_{C}$, for $k\log w = o(n)$, $w\in C$, $C\in\{{\tt poly, superpoly_\alpha, subexp_\alpha}\}$ and $\log w'=o(r) $, $r<k$  for any $w'\in C$.
\end{lemma}
\Beginproof
At first we use property, that was proved in \cite{agky14}:

\begin{equation}\label{peq2}
EQS_d\not\in{\bf NOBDD}_{2^{d/4}-1}.
 \end{equation}

Let us assume that there exits an $(k/r)$-NOBDD $P$ of width $w\in C$ such that $P$ computes $EQS_k$, hence $EQS_k\in  (k/r){\bf\mbox{-}NOBDD}_w$. Then  by  Theorem \ref{similate} we have $EQS_k\in  {\bf NOBDD}_{w^{2k/r-1}}$.

Therefore, the following statement is true: $w^{2k/r-1}=2^{(2k/r-1)\log w}<2^{(2k\log w)/r}<2^{k/8}$, because $\log w = o(r)$. We have  $EQS_k\not\in  {\bf NOBDD}_{2^{k/4-1}}$ due to statement (\ref{peq2}), hence $EQS_k\not\in  {\bf NOBDD}_{2^{k/8}}$. This is contradicts to the statement $EQS_k\in  {\bf NOBDD}_{w^{2k/r-1}}$. Hence $EQS_k\not\in (k/r)\bf{\mbox{-}NOBDD}_{W}$.
\Endproof

\section{Lower Bounds for Nondeterministic and Deterministic Ordered  Read-$k$-times  Branching Programs. Communication Complexity Technique}\label{s-lower-small}

We start with necessary definitions and notations.

 Let $\pi=(\{x_{j_1},\dots, x_{j_u}\}, \{x_{i_1},\dots, x_{i_v}\})=(X_A,X_B)$
 be a partition of the set $X$ into two parts $X_A$ and $X_B=X\backslash X_A$. Below we will use equivalent notations $f(X)$ and $f(X_A, X_B)$.

Let  $f|_\rho$ be a subfunction of $f$, where  $\rho$ is a mapping $\rho:X_A \to \{0,1\}^{|X_A|}$.
Function $f|_\rho$ is obtained from $f$ by applying $\rho$. We denote $N^\pi(f)$ to be number of different subfunctions with respect to partition $\pi$.

Let $\Theta(n)$ be the set of all permutations of $\{1,\dots,n\}$.
We say, that  partition $\pi$ agrees with permutation
$\theta=(j_1,\dots, j_n)\in \Theta(n)$, if for some $u$, $1<u<n$ the
following is right: $\pi=(\{x_{j_1},\dots,
x_{j_u}\},\{x_{j_{u+1}},\dots, x_{j_n}\})$. We denote $\Pi(\theta)$
a set of all partitions which agrees with $\theta$.

Let $ N^\theta(f)=   \max_{\pi\in \Pi(\theta)} N^\pi(f), \qquad
N(f)=\min_{\theta\in \Theta(n)}N^\theta(f). $ 

From paper \cite{ak13} we have the following lower bound for deterministic case.

\begin{theorem}\label{th-main1-small}
\cite{ak13} Let function $f(X)$ is computed  by  $k$-OBDD $P$ of  width $w$,
then $N(f) \leq w^{(k-1)w+1}. $
\end{theorem}

We can proof lower bound for nondeterministic case by the similar way. 
\begin{theorem}\label{th-main2-small}
Let function $f(X)$ is computed  by  $k$-NOBDD $P$ of  width $w$,
then $N(f) \leq 2^{w\big((k-1)w+1\big)}. $
\end{theorem}

We present the proof in the next section.

\subsection{The Proof of Theorem \ref{th-main2-small}}
 The proof of the Theorem is based on the representation of $k$-NOBDD computation process as a
communication protocol of  special form (Lemma \ref{l1}) and on the
description  of computation  process in matrix form (Lemma
\ref{l2}).

Now we define two party $(2k-1)$-round automata communication nondeterministic protocol that simulates $k$-NOBDD computation.

\begin{definition}[Nondeterministic Automata protocol]
Let $t\geq 1$ be an  odd integer, $l\ge 1$,
$\pi$ be a partition of the set $X$ of
variables.

We define  an $(\pi, t, l)$
automata communication nondeterministic protocol $R$ as follows:

$R$ is a two party  $t$-round communication protocol. Protocol $R$ uses the partition $\pi$ of variables $X$
among Alice (A) and Bob (B). Let $\nu=(\sigma,\gamma)$ be a
partition of the input $\nu\in\{0,1\}^n$ according to $\pi=(X_A,X_B)$. Player A always
starts the computation and Player B produces a result.

\begin{description}

\item[{\rm Round 1}]
Player $A$ generates the first set of messages $\mu^1\subset \{ 0,1\} ^{l}$
($\mu^1=\mu^1(\sigma)$), nondeterministically  chooses one of the messages $m^1\in\mu^1$ and  sends it to Player $B$.

 \item[\rm Round 2]
 Player $B$ generates the second set of messages $\mu^2\subset \{ 0,1\} ^{l}$ ($\mu^2=\mu^2(m^1,\gamma)$)), nondeterministically  chooses one of the messages $m^2\in\mu^2$ and sends it to Player $B$.

\item[\rm Round 3]
Player $A$  generates $\mu^3\subset \{ 0,1\} ^{l}$ ($\mu^3=\mu^3(m^2,\sigma)$)), nondeterministically  chooses one of the messages $m^3\in\mu^3$ and sends it to Player $B$.
\item[\rm Round 4]
Player $B$  generates $\mu^4\subset \{ 0,1\} ^{l}$ ($\mu^4=\mu^4(m^3,\sigma)$)), nondeterministically  chooses one of the messages $m^4\in\mu^4$ and sends it to Player $A$. etc.

...
\item[\rm Round $t$]
Player $B$ receives $m^t$ and produces a result of computation $0$ or $1$.
    \end{description}
%
The result $R(\nu)$ of
computation  $R$ on $\nu$ is $1$ if exists at least one sequence $(m^1,\dots m^t)$, such that result is $1$, and $R(\nu)=0$ otherwise.
Boolean function $f(X)$ is computed by protocol $R$ (presented by
$R$) if
$f(\nu)=R(\nu)$ for all $\nu\in\{0,1\}^{n}$.

\end{definition}

\begin{remark} ``Automata''  property for protocol $R$ means the following fact: set of
messages on current round $\mu^j$ depends only on input and previous
message $m^{j-1}$.
\end{remark}

We say that $(\pi, t, l)$ automata nondeterministic
protocol $R$ is agreed with $k$-NOBDD $P$  if $t=2k-1$ and  $\pi\in\Pi(\theta(P))$.

\begin{lemma}\label{l1}
 Let function $f$ be computed by $k$-NOBDD $P$ of width $w$. Then $f$ can be computed by $(\pi, 2k-1, \log w)$ automata nondeterministic protocol $R$ that agreed with program $P$
\end{lemma}
\Beginproof Let us construct $(\pi,
2k-1, \log w)$ automata nondeterministic protocol $R$ by $k$-NOBDD $P$ with the
partition  $\pi \in \Pi(\theta(P))$. Let $\pi=(X_A,X_B)$, input $\nu=(\sigma,\gamma)$ according to $\pi$ and the protocol $R$ have two players $A$ and $B$.

\begin{description}
        \item[{\rm Round 1}]Player $A$ emulates the first layers of program $P$ on levels which tests variables from $X_A$. Then program $P$ reaches the set of vertexes $Ver^1$ and forms the set of messages $\mu^1$, each of the messages is a binary encoding of numbers of vertexes from $Ver^1$. After that $A$ nondeterministically chooses $m_1$, $m_1\in\mu_1$ and sends it to Player $B$.  
        \item[\rm Round 2] 
	Player $B$ gets $m^2$ and emulates the first layer of program $P$ on levels which tests variables from $X_B$, starting from the vertex which number was encoded in $m^1$. Then program $P$ reaches the set of vertexes $Ver^2$ and forms set of messages $\mu^2$, each of the messages is a binary encoding of numbers of vertexes from $Ver^2$. After that Player $B$ nondeterministicaly chooses $m_2$, $m_2\in\mu_2$ and sends it to Player $A$.          
        
        \item[\rm Round 3] 
Player $A$ gets $m^3$ and emulates the second layer of program $P$ on levels which tests variables from $X_A$, starting from the vertex which number was encoded in $m^2$. Then program $P$ reaches the set of vertexes $Ver^3$ and forms set of messages $\mu^3$, each of the messages is a binary encoding of numbers of vertexes from $Ver^3$. After that Player $A$ nondeterministicaly chooses $m_3$, $m_3\in\mu_3$ and sends it to Player $B$.

\item[\rm Round 4]
Player $B$ gets $m^3$ and emulates the second layer of program $P$ on levels which tests variables from $X_B$, starting from the vertex which number was encoded in $m^3$. Then program $P$ reaches the set of vertexes $Ver^4$ and forms set of messages $\mu^4$, each of the messages is a binary encoding of numbers of vertexes from $Ver^4$. After that Player $B$ nondeterministicaly chooses $m_4$, $m_4\in\mu_4$ and sends it to Player $A$. 

etc.

\item[\rm Round $2k-1$] 

Player $B$ gets $m^{2k-1}$ and emulates the last layer of program $P$ on levels which tests variables from $X_B$, starting from the vertex which number was encoded in $m^{2k-1}$. Then program $P$ reaches one of the sink nodes and returns answer.

    \end{description}

If there exists a path from root node to $1$-sink node in program $P$ then the sequence of messages  $m_1,\dots,m^{2k-1}$ should also be in existence. In this case protocol $R$ returns $1$. In other case protocol $R$ returns $0$. Hence $R(\sigma, \gamma)=P(\sigma, \gamma)$.
 \Endproof

\begin{lemma}\label{l2}
Let Boolean function $f(X)$ be computed by $(\pi, t, l)$ automata nondeterministic communication protocol $R$ then
\[ N^\pi(f) \leq  2^{\alpha}, \]
for $\alpha=2^l\big((t+1)2^{l-1}+l\big)$.
\end{lemma}

\Beginproof The main idea of the proof of Lemma \ref{l2} is to put
protocol computation in a matrix form and to compute number of special
sub-matrices of protocol computation matrix. 

Firstly, let us describe matrix form protocol computation.

First of all, we define matrix $M_R(\sigma,\gamma)$ that represents a computation procedure of protocol $R$ on input $\nu= (\sigma,\gamma)$.

\begin{displaymath}
M_R(\sigma,\gamma)=\left(\begin{array}{c|c}
0 & M_R(\sigma) \\
\hline
M_R(\gamma) & 0
\end{array}\right).
\end{displaymath}

We define sub-matrices $M_R(\sigma)$ and $M_R(\gamma)$ as following:

\begin{displaymath}
M_R(\sigma)=
\left(\begin{array}{c|c|c|c|c|c}
0&M_R^{(1)}(\sigma)& 0 & \dots & 0 & 0\\
\hline
0&0 & M_R^{(2)}(\sigma)& \dots & 0 & 0\\
\hline

0&0& 0 & \dots&  M_R^{(k-2)}(\sigma)& 0\\
\hline
0&0& 0 & \dots& 0 & M_R^{(k-1)}(\sigma)
\end{array}\right).
\end{displaymath}

\begin{displaymath}
M_R(\gamma)=\left(\begin{array}{c|c|c|c}
M_R^{(1)}(\gamma) & 0 & \dots & 0\\
\hline
0 & M_R^{(2)}(\gamma) & \dots & 0\\
\hline
\vdots &\vdots & \ddots& \vdots\\
\hline
0& 0& \dots & M_R^{(k-1)}(\gamma)\\
\hline
0& 0& \dots &0
\end{array}\right).
\end{displaymath}

The  blocks $M_R^{(i)}(\sigma)$ and $M_R^{(i)}(\gamma)$ are  $t$ blocks of size $2^l\times 2^l$.  The block $M_R^{(i)}(\sigma)$ describes computation of round $2i+1$. And block $M_R^{(i)}(\gamma)$ describes computation of round $2i$. If Alice (Bob) receives a message $m$ and a message $m'$ from the set of messages $\mu$ then in $m$-th row, $m'$-th column of block we put $1$. Otherwise we put $0$.

Additionally we define vectors $p^0(\sigma)$ and $q(\gamma)$ that describe the first and the last rounds respectively. Vector $p^0(\sigma)=(p^0_1,\dots p^0_{(2k-1)2^l})$ defines set of messages $\mu^1$, which was formed on the first round of protocol $R$. Each element of vector corresponds to one of $M_R(\sigma,\gamma)$ matrix's line. If message $r$ belongs to $\mu_1$ then  $p^0_r=1$, other elements are $0$.

Vector $q(\gamma)=(q_1,\dots q_{(2k-1)2^l})$.  Each element of vector corresponds to one of $M_R(\sigma,\gamma)$ matrix's line. Vector $q(\gamma)=(0,\dots,0,q^{(2k-1)}(\gamma))$, where $q^{(2k-1)}(\gamma)=(q_1,\dots q_{2^l})$ such that

\begin{displaymath}
q_r = \left\{ \begin{array}{ll}
1, & \textrm{if Player $A$ sends message $r$ on Round $(2k-1)$}\\
&\textrm{and Player $B$ returns $1$,}\\
0, & \textrm{otherwise.}
\end{array} \right.
\end{displaymath}

Secondly, let us show that for any $\nu\in\{0,1\}^n$ we have following statement:
\[R(\nu)=P(\sigma, \gamma)=neq\Bigg(p^0(\sigma)\cdot \Big(M_R(\sigma,\gamma)\Big)^{t-1}q^T(\gamma),0\Bigg).\]

where $q^T$ is transposed $q$ and
\begin{displaymath}
neq(x,y) = \left\{ \begin{array}{ll}
1, & x\neq y,\\
0, & x=y.
\end{array} \right.
\end{displaymath}

Let vector $p^j=(p^j_1,\dots p^j_{t2^l})$ describes computation of protocol $R$ after $j$ rounds on input $\nu=(\sigma,\gamma)$.
\begin{displaymath}
p^j_r = \left\{ \begin{array}{ll}
z, & \textrm{where $z>0$, if $r$  corresponds to message from $\mu^{j+1}$,}\\
0, & \textrm{otherwise.}
\end{array} \right.
\end{displaymath}
It implies that we can compute $p^j$ in such a manner: $p^j=p^0(\sigma)\Big(M_P(\sigma,\gamma)\Big)^j$.

According to the definition of $q(\gamma)$ we have following fact: $p^{2k-2}\cdot q^T(\gamma)>0$ iff protocol $R$ returns $1$. Hence the forthcoming statement is true:

\[R(\sigma, \gamma)=neq \Bigg(p^0(\sigma)\cdot \Big(M_R(\sigma,\gamma) \Big)^{t-1}q^T(\gamma),0\Bigg).\]

According to this statement we can see that if any $\sigma,\sigma'\in\{0,1\}^{|X_A|}$ such that $M_R(\sigma)=M_R(\sigma')$ and $p^0(\sigma)=p^0(\sigma')$ then $R(\sigma, \gamma)=R(\sigma', \gamma)$. The proof is as follows:
\[R(\sigma, \gamma)=neq\Bigg(p^0(\sigma)\cdot \Big(M_R(\sigma,\gamma) \Big)^{t-1}q^T(\gamma)\Bigg)=\]\[=neq\Bigg(p^0(\sigma')\cdot \Big(M_R(\sigma',\gamma) \Big)^{t-1}q^T(\gamma)\Bigg)=R(\sigma', \gamma).\]

It is easy to see that any subfunction $f|_\rho(Y)$, for $Y\in\{0,1\}^{|X_B|}$ is computed by $R(\sigma, Y)$. Hence $N^{\pi}(f)$ cannot greats number of different protocols $R(\sigma, Y)$. And according to previous fact, a number of different protocols $R(\sigma, Y)$ cannot greats number of different pairs $(M_R(\sigma),p^0(\sigma'))$.

Using combinatoric arguments we get that  number of such pairs does not greats $2^{\alpha}$. Therefore we proved claim of Lemma.\Endproof

Lemmas \ref{l1}, \ref{l2} and according to definition of $N(f)$ we can prove Theorem \ref{th-main2-small}.

Note that $k$-OBDD is partial case of $k$-NOBDD and the proof of Theorem \ref{th-main1-small} is very similar. The one difference is following: set $\mu_i$ always contains only one element.

\subsection{Lower bound for Boolean Function $SAF_{k,w}$.}\label{saf1}

We consider Boolean function $SAF_{k,w}$, which was defined in \cite{k2015}.

\paragraph{Function $SAF_{k,w}$.}
Let us  define Boolean function $SAF_{k,w}(X)$. Informal, we divide a input into two parts, and each part into $w$ blocks. Each block has address and value. Function is iterative:
\begin{itemize}
\item {\bf Phase $1$.} We find block with address $0$ in the first part of input and compute the value of this block. That is the address of the block from the second part.
\item {\bf Phase $2$.} We take the block from the second part with computed address and compute the value of the block and henceforth we get a new address of a new block from the first part. 

...
\item {\bf Phase $2k-1$.} We find a block with a new address in second (first) part and check the value of this block. If value of the block greats $0$ then value of $SAF_{k,w}(X)$ is $1$, and $0$ otherwise.   
\end{itemize}
 If we do not find the block with sought address on any phase then the value of $SAF_{k,w}(X)$ is also $0$.

Formally, Boolean function $SAF_{k,w}(X):\{0,1\}^n\to \{0,1\}$ for integer $k=k(n)$ and  $w=w(n)$ such that
\begin{equation}
 2kw(2w + \lceil \log k \rceil + \lceil \log 2w \rceil)<n.\label{kw}
\end{equation}
We divide input variables to $2kw$ blocks. There are $\lceil
n/(2kw)\rceil =a$ variables in each block.  After that we divide
each block to
 {\em address} and {\em value} variables. First  $\lceil\log k\rceil + \lceil\log 2w\rceil$ variables of block are {\em address}
 and other $a-\lceil\log k\rceil + \lceil\log 2w\rceil=b$ variables of block are {\em value}.

We call $x^{p}_{0},\dots,x^{p}_{b-1}$ {\em value} variables of $p$-th block and  $y^{p}_{0},\dots,y^{p}_{\lceil\log k\rceil + \lceil\log 2w\rceil}$ are {\em address} variables, for $p\in\{0,\dots,2kw-1\}$.

Boolean function $SAF_{k,w}(X)$ is iterative  process  based on the definition of the following six functions.

Function $AdrK:\{0,1\}^n\times\{0,\dots,2kw-1\}\to \{0,\dots,k-1\}$
obtains firsts part of block's address. This block will be used only
in a step of iteration which number is computed using this function:

\[AdrK(X,p)=\sum_{j=0}^{\lceil\log k\rceil-1}y^{p}_{j}\cdot 2^{j} (\textrm{mod
}k).\]

Function $AdrW:\{0,1\}^n\times\{0,\dots,2kw-1\}\to \{0,\dots,2w-1\}$ obtains the second part of block's address. It is the address of a block within one step of iteration:

\[AdrW(X,p)=\sum_{j=0}^{\lceil\log 2w\rceil-1}y^{p}_{j+\lceil\log k\rceil}\cdot 2^{j} (\textrm{mod
}2w).\]

Function $Ind:\{0,1\}^n\times\{0,\dots,2w-1\}\times\{0,\dots,k-1\}\to \{0,\dots,2kw-1\}$ obtains a number of block by a number of step and address within this step of iteration:

\begin{displaymath}
Ind(X,i,t) = \left\{ \begin{array}{ll}
p, & \textrm{where $p$ is minimal number of block such that}\\
& \textrm{$AdrK(X,p)=t$ and $AdrW(X,p)=i$}, \\
-1, & \textrm{if there are no such $p$}.
\end{array} \right.
\end{displaymath}

Function $Val:\{0,1\}^n\times\{0,\dots,2w-1\}\times\{1,\dots,k\}\to \{-1,\dots,w-1\}$ obtains the value of a block that has an address $i$ within $t$-th step of iteration:

\begin{displaymath}
Val(X,i,t) = \left\{ \begin{array}{ll}
\sum_{j=0}^{b-1}x^{p}_{j} (\textrm{mod }w), & \textrm{where }p=Ind(X,i,t)\textrm{, for $p\geq 0$}, \\
-1, & \textrm{if }Ind(X,i,t)<0.
\end{array} \right.
\end{displaymath}

Two functions $Step_1$ and $Step_2$ obtain the value of $t$-th step of iteration. Function $Step_1:\{0,1\}^n\times\{0,\dots,k-1\}\to \{-1,w\dots,2w-1\}$ obtains the base for value of step of iteration:

\begin{displaymath}
Step_1(X,t) = \left\{ \begin{array}{ll}
-1, & \textrm{if }  Step_2(X,t-1)=-1, \\
0, & \textrm{if }  t=-1,\\
Val(X,Step_2(X,t-1),t) + w, & \textrm{otherwise}.
\end{array} \right.
\end{displaymath}

Function $Step_2:\{0,1\}^n\times\{0,\dots,k-1\}\to \{-1,\dots,w-1\}$ obtains the value of $t$-th step of iteration:

\begin{displaymath}
Step_2(X,t) = \left\{ \begin{array}{ll}
-1, & \textrm{if }  Step_1(X,t)=-1, \\
0, & \textrm{if }  t=-1\\
Val(X,Step_1(X,t),t), & \textrm{otherwise}.
\end{array} \right.
\end{displaymath}

Note that the address of current block is computed on the previous step.

The result of Boolean function $SAF_{k,w}(X)$ is computed by the  following way:

\begin{displaymath}
SAF_{k,w}(X) = \left\{ \begin{array}{ll}
0, & \textrm{if }  Step_2(X,k-1)\leq 0, \\
1, & \textrm{otherwise}.
\end{array} \right.
\end{displaymath}

Let set $C$ be one of the following sets:
\begin{itemize}

\item
${\tt const}=\{w:$ $w$ is constant$, w>20\}$,

\item
${\tt superpolylog}=\{w: w=O(\log^z n)$, $z=const\}$,

\item
${\tt sublinear}_\alpha=\{w: w=O(n^\alpha)\}$, $0<\alpha<0.5$.
\end{itemize}

\begin{lemma}\label{peq2_kobdd-small}
$SAF_{\lfloor k/2\rfloor,\lfloor (w-1)/3 \rfloor}\not\in (k/\delta){\bf-OBDD}_{C}$, for $kw(\log_2 w) =o(n)$, $w\in C$, $C\in\{{\tt const, superpolylog, sublinear_\alpha}\}$ and $k>4, w>20, \delta>\frac{48v\log_2 v}{w\log_2 w}$,  for any $v,w\in C$.
\end{lemma}
\Beginproof
We use property, that was proved in \cite{k2015}:

For integer $k=k(n)$, $w=w(n)$ and Boolean function $SAF_{k,w}$,
such that inequality (\ref{kw}) holds, the following statement is
true: $N(SAF_{k,w})\geq w^{(k-1)(w-2)}$.

Hence we can compute the following bound:
\[N(SAF_{\lfloor k/2\rfloor,\lfloor (w-1)/3 \rfloor})\geq\left\lfloor\frac{w-1}{3}\right\rfloor^{(\lfloor k/2 \rfloor-1)(\lfloor\frac{w-1}{3}\rfloor-2)}\geq\]
\[
\geq(w^{1/2})^{\frac{1}{2}(k-2)(\frac{w-1}{3} - 2)}=
w^{\frac{1}{12}(k-2)(w-7)}>w^{\frac{kw}{48}}.
\]

Let us compute the following rate: $N(SAF_{\lfloor k/2\rfloor,\lfloor (w-1)/3 \rfloor})/(v^{(k/\delta-1)v + 1})$ for $v\in W$.

\[\frac{N(SAF_{\lfloor k/2\rfloor,\lfloor (w-1)/3 \rfloor})}{v^{(k/\delta-1)v + 1}}\geq
\frac{w^{kw/48}}{v^{(k/\delta-1)v + 1}}
>\]
\[>
2^{kw(\log_2 w)/48-kv(\log_2 v)/\delta}
=2^{(\delta w\log_2 w - 48v\log_2 v)k/(48\delta)}>1
\]

Hence $N(SAF_{\lfloor k/2\rfloor,\lfloor (w-1)/3 \rfloor}) > (v^{(k/\delta-1)v + 1})$ for any $v\in W$. And by Theorem \ref{th-main1-small} we have  $SAF_{\lfloor k/2\rfloor,\lfloor (w-1)/3 \rfloor}\not\in \lfloor k/\delta\rfloor{\bf\mbox{-}OBDD}_{C}$.
\Endproof
\begin{lemma}\label{peq2_knobdd-small}
$SAF_{\lfloor k/2\rfloor,\lfloor (w-1)/3 \rfloor}\not\in \lfloor k/\delta\rfloor{\bf\mbox{-}NOBDD}_{C}$, for $kw^2 = o(n)$, $w\in C$, $C\in\{{\tt const, superpolylog, sublinear_\alpha}\}$ and $k>4, w>20, \delta>\frac{48v^2}{w\log_2 w}$,  for any $v,w\in C$.
\end{lemma}
\Beginproof
The proof is based on the proof of the previous Lemma.

Let us compute the following rate: $N(SAF_{\lfloor k/2\rfloor,\lfloor (w-1)/3 \rfloor})/(2^{v((k/\delta-1)v + 1}))$ for $v\in W$.

\[\frac{N(SAF_{\lfloor k/2\rfloor,\lfloor (w-1)/3 \rfloor})}{2^{v((k/\delta-1)v + 1)}}\geq
\frac{w^{kw/48}}{2^{v((k/\delta-1)v + 1)}}
>
2^{\frac{k}{48\delta}(\delta w\log_2 w - 48v^2)}>1.
\]

Hence $N(SAF_{\lfloor k/2\rfloor,\lfloor (w-1)/3 \rfloor}) > 2^{v((k/\delta-1)v + 1}$ for any $v\in W$. And by Theorem \ref{th-main1-small} we have  $ SAF_{\lfloor k/2\rfloor,\lfloor (w-1)/3 \rfloor}\not\in \lfloor k/\delta\rfloor{\bf\mbox{-} NOBDD}_{C}$.
\Endproof

\section{Lower Bounds for Probabilistic Ordered  Read-$k$-times  Branching Programs. Communication Complexity Technique.}\label {s-lower-p-small}
We use similar technique as for deterministic and nondeterministic models.

\begin{theorem}\label{th-main3-small}
Let function $f(X)$ be computed  by bounded error  $k$-POBDD $P$ of  width $w$,
then 
\[N(f)\leq   \left(C_1k(C_2 +\log_2{w} + \log_2{k})\right)^{(k+1)w^2}\]
for some constants $C_1$ and $C_2$.
\end{theorem}

We present the proof in the next section.

\subsection{The Proof of Theorem \ref{th-main3-small}}
 The proof of the Theorem is based on the representation of $k$-POBDD computation process as a
communication protocol of  special form (Lemma \ref{l1}) and on the
description  of computation  process in matrix form (Lemma
\ref{l2}).

Let us define two party $(2k-1)$-round automata communication probabilistic protocol that simulates $k$-POBDD computation.

\begin{definition}[Probabilistic Automata Protocol]
Let $t\geq 1$ be an  odd integer, $l\ge 1$,
$\pi$ be a partition of the set $X$ of
variables.

We define  an $(\pi, t, l)$
automata communication probabilistic protocol $R$ (shorter probabilistic $(\pi, t, l)$-protocol) as follows:

$R$ is a two party  $t$-round communication protocol. Protocol $R$ uses the partition $\pi$ of variables $X$
among Alice (A) and Bob (B). Let $\nu=(\sigma,\gamma)$ be a
partition of the input $\nu\in\{0,1\}^n$ according to $\pi=(X_A,X_B)$. Player A always
starts the computation and Player B produces a result.

\begin{description}

\item[{\rm Round 1}]
Player $A$ generates the first set of messages $\mu^1\subset \{ 0,1\} ^{l}$
($\mu^1=\mu^1(\sigma)$), using probabilistic mechanism  chooses one of the messages $m^1\in\mu^1$ and  sends it to Player $B$.

 \item[\rm Round 2]
 Player $B$ generates the second set of messages $\mu^2\subset \{ 0,1\} ^{l}$ ($\mu^2=\mu^2(m^1,\gamma)$)), using probabilistic mechanism  chooses one of the messages $m^2\in\mu^2$ and sends it to Player $B$.

\item[\rm Round 3]
Player $A$  generates $\mu^3\subset \{ 0,1\} ^{l}$ ($\mu^3=\mu^3(m^2,\sigma)$)), using probabilistic mechanism  chooses one of the messages $m^3\in\mu^3$ and sends it to Player $B$.
\item[\rm Round 4]
Player $B$  generates $\mu^4\subset \{ 0,1\} ^{l}$ ($\mu^4=\mu^4(m^3,\sigma)$)), using probabilistic mechanism  chooses one of the messages $m^4\in\mu^4$ and sends it to Player $A$. etc.

...
\item[\rm Round $t$]
Player $B$ receives $m^t$ and produces a result of computation $0$ or $1$.
    \end{description}
%
The result $R(\nu)$ of
computation  $R$ on $\nu$ is $1$ if probability of $1$-result greats $0.5+\delta$ for some $\delta>0$, and if probability of $0$-result greats $0.5+\delta$ then $R(\nu)=0$.
Boolean function $f(X)$ is computed by protocol $R$ (presented by
$R$) if
$f(\nu)=R(\nu)$ for all $\nu\in\{0,1\}^{n}$.

\end{definition}

We say that probabilistic $(\pi, t, l)$-protocol $R$ is agreed with $k$-POBDD $P$  if $t=2k-1$ and  $\pi\in\Pi(\theta(P))$.

\begin{lemma}\label{pl1}
 Let function $f$ be computed by $k$-POBDD $P$ of width $w$. Then $f$ can be computed by $(\pi, 2k-1, \log w)$ automata probabilistic protocol $R$ that agreed with program $P$
\end{lemma}
\Beginproof We can proof it by the same way as for Lemma \ref{l1}.
 \Endproof

\begin{lemma}\label{pl2}
Let Boolean function $f(X)$ be computed by probabilistic $(\pi, t, l)$-protocol $R$ then
\[N^{\pi}(f)\leq   \left(C_1t(C_2 + l + \log_2{(t+1)})\right)^{(t+3)2^{2l-1}}\]
for some constants $C_1$ and $C_2$.

\end{lemma}

The proof of the lemma is presented in the next section.

We can prove Theorem \ref{th-main3-small} due to Lemmas \ref{pl1}, \ref{pl2} and definition of $N(f)$ .

\subsection{The Proof of Lemma \ref{pl2}}
 The main idea of the proof of Lemma \ref{pl2} is similar to proof of Lemma \ref{l2}. 

Firstly, we consider similar matrices  $M_R(\sigma,\gamma)$, $M_R(\sigma)$ and $M_R(\gamma)$. $M_R(\sigma)$ consist from $M_R^{(i)}(\sigma)$ and $M_R(\gamma)$ consist from $M_R^{(i)}(\gamma)$.

The  blocks $M_R^{(i)}(\sigma)$ and $M_R^{(i)}(\gamma)$ are  $t$ blocks of size $2^l\times 2^l$.  The block $M_R^{(i)}(\sigma)$ describes computation of round $2i+1$. And block $M_R^{(i)}(\gamma)$ describes computation of round $2i$. If Alice (Bob) receives a message $m$ and  then sends message $m'$ with probability $pr$ then in $m$-th row, $m'$-th column of block we put $pr$.

Similar to nondeterministic case we define vectors $p^0(\sigma)$ and $q(\gamma)$ that describe the first and the last rounds respectively. Vector $p^0(\sigma)=(p^0_1,\dots p^0_{(2k-1)2^l})$ defines set of messages $\mu^1$, which was formed on the first round of protocol $R$. Each element of vector corresponds to one of $M_R(\sigma,\gamma)$ matrix's line. If message $r$ belongs to $\mu_1$ then  $p^0_r=Pr\{r\}$, where $Pr\{r\}$ is probability of sending $r$ message.

Vector $q(\gamma)=(q_1,\dots q_{(2k-1)2^l})$.  Each element of vector corresponds to one of $M_R(\sigma,\gamma)$ matrix's line. Vector $q(\gamma)=(0,\dots,0,q^{(2k-1)}(\gamma))$, where $q^{(2k-1)}(\gamma)=(q_1,\dots q_{2^l})$. $q_r$ is probability of accepting input if Player $A$ sends message $r$ on Round $(2k-1)$.

Secondly, let us show that for any $\nu\in\{0,1\}^n$ we have following statement:
\[Pr\{R\mbox{ reaches }1\mbox{ on }\nu\}=p^0(\sigma)\cdot \Big(M_R(\sigma,\gamma)\Big)^{t-1}q^T(\gamma).\]

Let vector $p^j=(p^j_1,\dots p^j_{t2^l})$ describes computation of protocol $R$ after $j$ rounds on input $\nu=(\sigma,\gamma)$. $p^j_r$ is probability of sending message $r$ on Round $j+1$.

It implies that we can compute $p^j$ in such a manner: $p^j=p^0(\sigma)\Big(M_P(\sigma,\gamma)\Big)^j$.

According to the definition of $q(\gamma)$ we can see that $p^{2k-2}\cdot q^T(\gamma)$  is probability of $1$-result. Hence the forthcoming statement is true:

\[Pr\{R\mbox{ reaches }1\mbox{ on }(\sigma, \gamma)\}=p^0(\sigma)\cdot \Big(M_R(\sigma,\gamma) \Big)^{t-1}q^T(\gamma).\]

Note, that in probabilistic case we cannot discuss just equality of matrices $M_R(\sigma)$ and $M_R(\sigma')$; vectors $p^0(\sigma)$ and $p^0(\sigma')$. The reason is following: if they have vary small difference then probability of $1$-result can be enough small for accepting input in both cases. 

Let us discuss measure of matrix closeness.

\subsubsection{Measure of Matrix Closeness}

Let $\beta \geq 1$. Two numbers $p$ and $p'$ are said to be $\beta$-close if either
\begin{itemize}
\item $p = p' = 0$ or
\item $p \geq 0$, $p' \geq 0$, and $\beta ^{-1} \leq p / p' \leq \beta$.
\end{itemize} 

Two $r \times r$ matrices $M$ and $M'$, such that $s_{i,j}$ are elements of $M$ and $s'_{i,j}$ are elements of $M'$, are $\beta$-close if $s_{i,j}$ and $s'_{i,j}$ are $\beta$-close for each $1\leq i,j \leq r$. We can say about $\beta$-closeness of vectors by similar way.

Let us discuss some properties of this metric.

\begin{property}\label{meraSum}
 If $a$ and $b$ are $\beta$-close; $c$ and $d$ are $\beta$-close  then $(a+c)$ and $(b+d)$ are $\beta$-close. 
\end{property}
 \Beginproof
Firstly, let us prove that $(a+c)/(b+d)\geq 1/\beta$:
\[\frac{a+c}{b+d}\geq\frac{\frac{1}{\beta}b +\frac{1}{\beta}d}{b+d}=1/\beta\]
Secondly, let us prove that $(a+c)/(b+d)\leq\beta$:
\[\frac{a+c}{b+d}\leq\frac{\beta b +\beta d}{b+d}=\beta.\]

If $a=0$ and $b=0$, then $a+c=c$ and  $b+d=d$, therefore $(a+c)$ and $(b+d)$ are $\beta$-close.

If $c=0$ and $d=0$, then $a+c=a$ and  $b+d=c$, therefore $(a+c)$ and $(b+d)$ are $\beta$-close.
 \Endproof
 
 \begin{property}\label{meraProd}
If $a$ and $b$ are $\beta$-close; $c$ and $d$ are $\beta$-close, then $ac$ and $bd$ are $\beta'\cdot\beta$-close
\end{property}
 \Beginproof
If $a,b,c,d\neq 0$, then claim follows from definition of $\beta$-closeness.

If $a=0$ and $b=0$, then $ac=0$ and $bd=0$, therefore $ac$ and $bd$ are $\beta'\beta$-close.

If $c=0$ and $d=0$, then $ac=0$ and $bd=0$, therefore $ac$ and $bd$ are $\beta'\beta$-close.
 \Endproof

\begin{property}\label{mulScalar}
 If $a$ and $b$ $\beta$-close, then $d\cdot a$ and $d\cdot b$ are $\beta$-close for any $d\geq 0$.
\end{property}
 \Beginproof
If $a,b,d\neq 0$, then claim follows from definition of $\beta$-closeness.

If $a=0$ and $b=0$ or $d=0$, then $d\cdot a=0$ and  $d\cdot b=0$, therefore $d\cdot a$ and $d\cdot b$ also $\beta$-close.
 \Endproof

 \begin{property}{\label{twoMatrixLem}}
 If two $r \times r$ matrices $B_1=[c_{ij}]$ and $B_2=[c'_{ij}]$ are $\beta$-close, then two matrices $(B_1)^z$ and $(B_2)^z$ are $\beta^z$-close
 \end{property}  
 \Beginproof
Elements of matrices $(B_1)^z=[\tilde{c}_{ij}]$ and $(B_2)^z=[\tilde{c}'_{ij}]$ such that: 
 \[
 \tilde{c}_{ij}=\sum_{q_1,\dots,q_{z-1}=1}^{r}{c_{iq_1}c_{q_1q_2}c_{q_2q_3}\dots c_{q_{z-1}j}}=\sum_{q_1,\dots,q_{z-1}=1}^{r}C_{iq_1q_2\dots q_{z-1}j},
 \]
 \[
 \tilde{c}'_{ij}=\sum_{q_1,\dots,q_{z-1}=1}^{r}{c'_{iq_1}c'_{q_1q_2}c'_{q_2q_3}\dots c'_{q_{z-1}j}}=\sum_{q_1,\dots,q_{l-1}=1}^{r}C'_{iq_1q_2\dots q_{z-1}j}.
 \]
 
Let us discuss $\beta$-closeness of $\tilde{c}_{ij}$ and $\tilde{c}'_{ij}$.
 
 $C_{iq_1q_2\dots q_{z-1}j}$ and $C'_{iq_1q_2\dots q_{z-1}j}$ are $\beta^z$-close due to Property \ref{meraProd}. Additionally, $\sum_{q_1,\dots,q_{z-1}=1}^{r}C_{iq_1q_2\dots q_{z-1}j}$ and ${\sum_{q_1,\dots,q_{z-1}=1}^{r}C_{iq_1q_2\dots q_{z-1}j}}$ are $\beta^z$-close due to Property \ref{meraSum}. Hence matrices $B_1^z$  and $B_2^z$ are $\beta^z$-close.
 \Endproof

Let us discuss how $\beta$-closeness of matrices $M(\sigma, \gamma)$ and $M(\sigma', \gamma)$ and vectors  $p^0(\sigma)$ and $p^0(\sigma)$ affects to probability of $1$-result for these two inputs. 

\begin{lemma}\label{bClosenessOfAcceptPr}
 Let inputs $\nu=(\sigma, \gamma)$ and $\nu'=(\sigma',\gamma)$ be such that  $M(\sigma,\gamma)$ and $M(\sigma',\gamma)$ are $\beta$-close and $p^0(\sigma)$ and $p^0(\sigma)$ are $\beta$-close. Following fact is true: probabilities of $1$-result  $Pr\{R\mbox{ reaches }1\mbox{ on }(\sigma, \gamma)\}$ and $Pr\{R\mbox{ reaches }1\mbox{ on }(\sigma', \gamma)\}$ are $\beta^{2k-1}$-close
 \end{lemma}  
 \Beginproof
 Firstly, matrices $M(\sigma,\gamma)^{2k-2}$ and $M(\sigma',\gamma)^{2k-2}$ are $\beta^{2k-2}$-close due to Property \ref{twoMatrixLem}.
 
 Secondly, we can prove that $p^0(M(\sigma,\gamma)^{2k-2})$ and $p^0(M(\sigma',\gamma)^{2k-2})$ are $\beta^{2k-1}$-close by the way similar to the proof of Property \ref{twoMatrixLem}.
 
 Finally, $p^0(M(\sigma,\gamma)^{2k-2})q^T(\gamma)$ and $p^0(M(\sigma',\gamma)^{2k-2})q^T(\gamma)$ are $\beta^{2k-1}$-close due to previous fact and Properties  \ref{mulScalar} and \ref{meraProd}. Therefore, $Pr\{R\mbox{ reaches }1\mbox{ on }(\sigma, \gamma)\}$ and $Pr\{R\mbox{ reaches }1\mbox{ on }(\sigma', \gamma)\}$ are $\beta^{2k-1}$-close.
 \Endproof

 Let us define {\em weak} protocol $R'$ for protocol $R$.
 
 \begin{definition}
 Let probabilistic $(\pi,t,l)$-protocol $R$ computes Boolean function $f$ with bounded error $\delta$.  Week probabilistic $(\pi,t,l)$-protocol $R'$ is protocol which was obtained from protocol $R$ by the following way:
 
 For any input $\nu=(\sigma,\gamma)$ 
 \begin{itemize}
 \item vectors $p^0(\sigma)$ and $q(\gamma)$ is the same as for $R$;
 \item Let $M'(\sigma,\gamma)=[s'_{i,j}]$ corresponds to $R'$ and $M(\sigma,\gamma)=[s_{i,j}]$ corresponds to $R$ then 
 \begin{displaymath}
s'_{ij} = \left\{ \begin{array}{ll}
s_{ij}, & s_{ij}\geq\frac{\delta}{16}k^{-4}2^{-5l},\\
0, & \textrm{otherwise.}
\end{array} \right.
\end{displaymath}
for $t=2k-1$.
 \end{itemize} 
   
 \end{definition}

{\em Weak} protocol have the following property:

\begin{lemma}\label{cpp_small_elems}
If protocol $R$ computes Boolean function $f$ with bounded error $\delta$ then weak protocol $R'$ computes $f$ with bounded error $\delta/2$.
\end{lemma}  
\Beginproof
 Probability of $1$-result is  $p_0(M_P(\sigma,\gamma)^{2k-2})q^T$, the upper bound for the probability is $2^{2l}\tilde{s}_{i_0j_0}$, where $\tilde{s}_{i_0j_0}$ is one of  $M_P(\sigma,\gamma)^{2k-2}$ matrix elements.
 
Let $\tilde{s}_{i_0j_0}=h+h(s_{ij})$, where $h$ is sum of all	summands which does not contains $s_{ij}$ and $h(s_{ij})$ is sum of all	summands which contains $s_{ij}$.
 
 \[h(s_{ij})\leq \sum_{q_1,\dots,q_{2k-2}=1}^{k2^l}{s_{i_0q_1}s_{ij}s_{q_2q_3}\dots s_{q_{2k-2}j_0}} +
 \sum_{q_1,\dots,q_{2k-2}=1}^{k2^l}{s_{i_0q_1}s_{q_1q_2}s_{ij}\dots s_{q_{2k-2}j_0}} +\]
 \[+\dots+\sum_{q_1,\dots,q_{2k-2}=1}^{k2^l}{s_{i_0q_1}s_{q_1q_2}s_{q_2q_3}\dots s_{ij}s_{q_{2k-3}j_0}} +h'(s_{ij})+h''(s_{ij}), \]
where

\begin{displaymath}
h'(s_{ij}) = \left\{ \begin{array}{ll}
\sum_{q_1,\dots,q_{2k-2}=1}^{k2^l}{s_{ij}s_{q_1q_2}s_{q_2q_3}\dots s_{q_{2k-2}j_0}}, & i=i_0,\\
0, & \textrm{otherwise.}
\end{array} \right.
\end{displaymath}

\begin{displaymath}
h''(s_{ij}) = \left\{ \begin{array}{ll}
\sum_{q_1,\dots,q_{2k-2}=1}^{k2^l}{s_{i_0q_1}s_{q_1q_2}s_{q_2q_3}\dots s_{q_{2k-3}q_{2k-2}} s_{ij}}, & j=j_0,\\
0, & \textrm{otherwise.}
\end{array} \right.
\end{displaymath}

Let us consider the following sums:
\[\sum_{q_1,\dots,q_{2k-2}=1}^{k2^l}{\big((s_{i_0q_1}s_{q_1q_2}\dots s_{q_{r-1}q_r})s_{ij}(s_{q_{r+1}q_{r+2}}\dots s_{q_{2k-2}j_0})\big)}=\]\[=
\left(\sum_{q_1,\dots,q_{r}=1}^{k2^l}{s_{i_0q_1}s_{q_1q_2}\dots s_{q_{r-1}q_r}}\right)s_{ij}\left(\sum_{q_{r+1},\dots,q_{2k-2}=1}^{k2^l}{s_{q_{r+1}q_{r+2}}\dots s_{q_{2k-2}j_0}}\right)=\]\[=
\left(\sum_{q_{r}=1}^{k2^l}{\sum_{q_1,\dots,q_{r-1}=1}^{k2^l}{s_{i_0q_1}s_{q_1q_2}\dots s_{q_{r-1}q_r}}}\right)s_{ij}\left(\sum_{q_{r+1}=1}^{k2^l}{\sum_{q_{r+2},\dots,q_{2k-2}=1}^{k2^l}{s_{q_{r+1}q_{r+2}}\dots s_{q_{2k-2}j_0}}}\right)\leq\]
\[\leq \left(\sum_{q_{r}=1}^{k2^l}{s_{i_0q_r}^{[r]}}\right)s_{ij}\left(\sum_{q_{r+1}=1}^{k2^l}{s_{q_{r+1}j_0}^{[2k-2-r]}}\right)\leq \left(\sum_{q_{r}=1}^{k2^l}{1}\right)s_{ij}\cdot 1\leq
k 2^{l}s_{ij},\]
where $s_{ij}^{[z]}$ is element of matrix $\left(M_P(\sigma,\gamma)\right)^z$.
This inequality right due to $M_P(\sigma,\gamma)$ is stochastic and hence $\left(M_P(\sigma,\gamma)\right)^r$ and $\left(M_P(\sigma,\gamma)\right)^{2k-2-r}$ also stochastic, therefore sum of all elements of a row of the matrix less or equal $1$.

Let us continue to estimate $h(s_{ij})$. According to above statements we have:
\[h(s_{ij})\leq (2k-1)k 2^ls_{ij}\leq 2k^22^{l}s_{ij}\leq\frac{\delta}{8}k^{-2}2^{-4l}.\]

Number of elements which becomes $0$ in $M'(\sigma,\gamma)$ less or equals to number of all elements: $(2k-1)^2\left(2^l\right)^2$. Therefore all this elements give summed in $1$-result probability less or equals to 
    \[(2k-1)^22^{2l}h(s_{ij})\leq(2k-1)^22^{2l}\frac{\delta}{8}k^{-2}2^{-4l}\leq \frac{\delta}{2}2^{-2l}.\]

Probability of $R'$ protocol's $1$-result for input  $\sigma,\gamma$ is   $Pr\{R'\mbox{ reaches }1\mbox{ on }(\sigma, \gamma)\}\geq Pr\{R\mbox{ reaches }1\mbox{ on }(\sigma, \gamma)\} - 2^{2l}\cdot(2k-1)2^{2l}h(s_{ij})2^l\geq Pr\{R(\sigma,\gamma)\mbox{ reaches }1\} -\delta/2$.  Hence we can say that
$|Pr\{R'\mbox{ reaches }1\mbox{ on }(\sigma, \gamma)\}-Pr\{R\mbox{ reaches }1\mbox{ on }(\sigma, \gamma)\}|\leq \delta/2$.
\Endproof

Let us discuss closeness of inputs according to probabilistic protocol.

\begin{lemma}\label{cpp_eps_eq}
Let probabilistic $(\pi,t,l)$-protocol $R$ computes Boolean function $f$ with bounded error $\delta$, inputs $\sigma$ and $\sigma'$ are $\beta$-equivalent, it means corresponding $M(\sigma)$ and $M(\sigma')$ and $p^0(\sigma)$ and $p^0(\sigma')$ are $\beta$-close, for any $\gamma\in\{0,1\}^{|X_B|}$ and $\beta=\left(( 0.5-\delta/2)/(0.5+\delta/2)\right)^{1/(2k-1)}$. Then $R$ returns $1$ for input $\nu'=(\sigma',\gamma)$ with bounded error $\delta/2$ iff $R$ returns $1$ for input $\nu=(\sigma,\gamma)$ with bounded error $\delta$ for any $\gamma\in\{0,1\}^{|X_B|}$. And same situation for $0$-result.  
\end{lemma}
\Beginproof
 Let $p=Pr(R(\sigma,\gamma)=1)$ be probability of $1$-result for $\nu=(\sigma,\gamma)$ and $p'=Pr\{R\mbox{ reaches }1\mbox{ on }(\sigma', \gamma)\}$ be probability of $1$-result for $\nu'=(\sigma',\gamma)$. 

Assume that $R(\sigma,\gamma)=1$, it means that $p>0.5+\delta$. Probabilities $p$ and $p'$ are $\beta^{2k-1}$-close due to Lemma \ref{bClosenessOfAcceptPr}. Hence we have:

\[
p'> \beta^{-(2k-1)}p>\beta^{-(2k-1)}(0.5+\delta)=\frac{0.5-\delta/2}{0.5-\delta}(0.5+\delta)>0.5+\delta/2.
\]

Assume that $R(\sigma,\gamma)=0$, it means that $p<0.5-\delta$. Hence we have:
\[
p'<\beta^{2k-1}p< \beta^{2k-1}(0.5-\delta)=\frac{0.5-\delta/2}{0.5-\delta}(0.5-\delta)=0.5-\delta/2.
\]

Summaries we have following two facts:
\begin{itemize}
\item if $p>0.5+\delta$ then $p'>0.5+\delta/2$;
\item if $p<0.5-\delta$ then $p'<0.5-\delta/2$.
\end{itemize}
Therefore if $R$ returns $1$ on $(\sigma,\gamma)$ with bounded error $\delta$ then $R$ returns $1$ on $(\sigma',\gamma)$ with bounded error $\delta/2$; and vise verse if $R$ returns $0$ on $(\sigma,\gamma)$ with bounded error $\delta$ then $R$ returns $0$ on $(\sigma',\gamma)$ with bounded error $\delta/2$.
\Endproof

\subsubsection{Final Phase of The Proof of Lemma \ref{pl2}}
Let protocol $R$ computes Boolean function $f$ with bounded error $2\delta$. Then We can build {\em weak} protocol $R'$ which computes $f$ with bounded error $\delta$ due to Lemma \ref{cpp_small_elems}.

Let us proof that number of subfunctions $N^\pi(f)$ less or equals number of non $\beta$-equivalent $\sigma$s which are corresponds to Protocol $R'$ and error $\delta$.

Assume that  $N^\pi(f)$ greats number of non $\beta$-equivalent $\sigma$s which are corresponds to protocol $R'$ and error $\delta$. Then due to Pigeonhole principal there are two inputs $\sigma$ and $\sigma'$ and corresponding mappings $\rho$ and $\rho'$ such that $f|_\rho(X_B)\neq f|_\rho'(X_B)$, but $\sigma$ and $\sigma'$ are $\beta$-equivalent. It means there are $\gamma\in\{0,1\}^{|X_A|}$ such that $f|_\rho(\gamma)\neq f|_\rho'(\gamma)$, but $R(\sigma,\gamma)=R(\sigma',\gamma)$ with bounded error $\delta/2$. This is contradiction.

Let us compute number of different non $\beta$-equivalent $\sigma$s, it means number of non $\beta$-close matrices $M(\sigma)$ multiply number of non $\beta$-close matrices $p^0(\sigma)$.

We remember that $R'$ is weak protocol. Hence we have following bound for number of non $\beta$-close matrices $M(\sigma)$:
\[\left(\frac{1-\log_2{((\delta/8 )k^{-3}2^{-5l})}}{\log \beta} \right)^{k2^{2l}}\leq
\left(\frac{2+5(l+\log_2{k})-(\log_2{\delta}-3)}{\log \beta}\right)^{k2^{2l}}\leq\]\[\leq
\left(\frac{5}{\log{\beta}}(1 +l + \log_2{k}-0.2\log_2{\delta})\right)^{k2^{2l}}.\] 

Additionally we have following bound for number of non $\beta$-close matrices $p^0(\sigma)$:
\[\left(\frac{5}{\log{\beta}}(1 +l + \log_2{k}-0.2\log_2{\delta})\right)^{2^{l}}.\] 
Therefore
\[N^{\pi}(f)\leq  \left(\frac{5t}{\log_2{\frac{0.5-\delta/2}{0.5 - \delta}}}(l + \log_2{(t+1)}-0.2\log_2{\delta})\right)^{(t+3)2^{2l-1}}.\]

Then we have

\[N^{\pi}(f)\leq   \left(C_1t(C_2 + l + \log_2{(t+1)})\right)^{(t+3)2^{2l-1}}\]
for some constants $C_1$ and $C_2$. It means that for any $\theta\in \Theta(n)$ we have
\[N^{\theta}(f)\leq   \left(C_1t(C_2 +l + \log_2{(t+1)})\right)^{(t+3)2^{2l-1}}\]
and
\[N(f)\leq   \left(C_1t(C_2 +l + \log_2{(t+1)})\right)^{(t+3)2^{2l-1}}.\]

\subsection{Lower bound for Boolean Function $SAF_{k,w}$.}\label{saf2}
\begin{lemma}\label{peq2_kpobdd-small}
$SAF_{\lfloor k/2\rfloor,\lfloor (w-1)/3 \rfloor}\not\in (k/\psi){\bf\mbox{-}POBDD}_{C}$, for $kw^2\log(k(\log w + \log k))=o(n)$, $w\in C$, $C\in\{{\tt const, superpolylog, sublinear_\alpha}\}$ and $k>4, w>20, v^2\log (k(\log k + \log v))=o(\psi)$,  for any $v,w\in C$.
\end{lemma}
\Beginproof
In the proof of Lemma \ref{peq2_kobdd-small} we showed that
$N(SAF_{\lfloor k/2\rfloor,\lfloor (w-1)/3 \rfloor})>w^{\frac{kw}{48}}
$.

Let us compute the following rate for $v\in W$.

\[\frac{N(SAF_{\lfloor k/2\rfloor,\lfloor (w-1)/3 \rfloor})}{\left(C_1(k/\psi)
(\log_2 v +\log_2{(k/\psi)}+C_2)\right)^{(k/\psi+1)v^{2}}}
>\]
\[>
\frac{w^{\frac{kw}{48}}}{\left((C_3k/\psi)
(\log_2 v +\log_2{(k/\psi)})\right)^{(k/\psi+1)v^{2}}}>\]
\[=
2^{\frac{k}{48\psi}\left(\psi w\log_2 w - 96 v^{2}\log_2\left(C_3 k/\psi
(\log_2 v+\log_2{(k/\psi)})\right)\right)}>1\]

Hence $N(SAF_{\lfloor k/2\rfloor,\lfloor (w-1)/3 \rfloor}) > \left(C_1(k/\psi)
(\log_2 v +\log_2{(k/\psi)}+C_2)\right)^{(k/\psi+1)v^{2}})$ for any $v\in W$. And by Theorem \ref{th-main1-small} we have  $SAF_{\lfloor k/2\rfloor,\lfloor (w-1)/3 \rfloor}\not\in \lfloor k/\psi\rfloor{\bf\mbox{-}OBDD}_{C}$
\Endproof

\section{Hierarchy Theorems}\label{sec:hierarchy}
\subsection{Results for Polynomial, Superpolynomial and Subexponential Width}

\begin{theorem}\label{th-hi}
 For $k=k(n)$ and $w=w(n)$ such that $k\log w = o(n)$, $w\in C$ for set $C\in\{{\tt poly, superpoly_\alpha, subexp_\alpha}\}$, the following inclusion is true:
\[
\left\lfloor k/r\right\rfloor \bf{\mbox{-}NOBDD}_{C}\subsetneq \bf{\mbox{-}NOBDD}_{C}
\]
for $\log w'=o(r) $, $r<k$  for any $w'\in C$.
\end{theorem}
\Beginproof
Clearly we have the situation when $\left\lfloor k/r\right\rfloor\bf{\mbox{-}NOBDD}_{C}\subseteq k\bf{\mbox{-}NOBDD}_{C}$. In order to prove proper inclusion we use properties of the Boolean function $EQS_k$.

\begin{lemma}\label{peq1}
There is $k$-OBDD $P$ of width $k^2 + 6k+2$ which computes Boolean function $EQS_k$.
\end{lemma}
\Beginproof
 Here we present a construction of $k$-OBDD $P$ with all necessary parameters  for $EQS_k$. $P$ tests variables $x_1,\dots,x_n$ in natural order. Let $\nu\in\{0,1\}^n$ be an input. The description of computation is following.
 
Note that Branching program can be decried using ``if-then-goto-else-goto'' instructions. And we will employ this representation for describe $P$.

The program $P$ on the layer $i$ compare $i$-th bit of $\alpha(\nu)$ and $\beta(\nu)$. If $i$-th bits are equal, then program goes to the $i+1$ layer for comparing $(i+1)$-st bits, and rejects input otherwise. In the last layer we have equality of $k$-th bits hence
$\alpha(\nu)=\beta(\nu)$ and $P$ accepts input.

Let us describe layer $i$ of program $P$. There are four groups of nodes in each level:

The first group contains $i^2$ nodes and the program $P$ riches these nodes when $i$-th bit form $\alpha(\nu)$ and $i$-th bit form $\beta(\nu)$ are not tested. We associate each node with the pair $(j,r)$, for $j,r\in\{0,\dots,i-1\}$, which means that the program $P$ has  already tested $j$ {\em value} bits from $\alpha(\nu)$  and $r$ {\em value} bits from $\beta(\nu)$. We denote these nodes as $node(i,z,1,j,r)$ on level which reads $x_z$.

The second group contains $3i$ nodes and $P$ riches these nodes when $i$-th bit form $\alpha(\nu)$ is not tested, but $i$-th bit form $\beta(\nu)$ is already tested. Each node of the group is associated with pair $(q,j)$, for $q\in\{-1, 0,1\}, j\in\{1,\dots, i\}$, which means $i$-th {\em value} bit from $\beta(\nu)$ equals $q$ and the program has already read $j$ {\em value} bits from $\alpha(\nu)$. The situation when $q=-1$ can be only on the level which reads  {\em value} bits. It means that current {\em value} is  $i$-th bit form $\beta(\nu)$ and the program has to store it. We denote these nodes as $node(i,z,2,q,j)$ on the level which reads $x_z$.

The third group contains $3i$ nodes and $P$ reaches these nodes when $i$-th bit form $\beta(\nu)$ is not read, but $i$-th bit form $\alpha(\nu)$ is read. Each node of the group is associated with pair $(q,r)$, for $q\in\{-1, 0,1\}, j\in\{1,\dots, i\}$, which means that $i$-th {\em value} bit from $\alpha(\nu)$ equals $q$ and the program has already read $r$ {\em value} bits from $\beta(\nu)$. The situation when $q=-1$ can be only on the level which reads  {\em value} bits, and it means that current {\em value} is  $i$-th bit form $\alpha(\nu)$ and the program should store it. We denote this nodes as $node(i,z,3,q,r)$ on level which reads $x_z$.

The fourth group contains two nodes: $equals$, which means $i$-th bits of $\alpha(\nu)$ and $\beta(\nu)$ equals;  $reject$, which means $i$-th bits of $\alpha(\nu)$ and $\beta(\nu)$ do not equals. We denote this nodes as $node(i,z,equals)$ and $node(i,z,reject)$ on the level which reads $x_z$.

Let us describe the code of the layer $i$, level $z$ (in code we use $x[z]$ for $x_z$), for $i\in\{1,\dots,k\},z\in\{1,\dots,n-1\}$.

If $x_z$ is {\em marker} bit then we have following program:

\begin{itemize}
\item For $j,r\in\{1,\dots,i-2\}$ we have:
\begin{verbatim}
node(i,z,1,j,r):
if x[z]=0 then goto node(i,z+1,1,j+1,r) 
else goto node(i,z+1,1,j,r+1);
\end{verbatim}
\item For $r\in\{1,\dots,i-2\}$ we have:
\begin{verbatim}
node(i,z,1,i-1,r): 
	if x[z]=0 then goto node(i,z+1,2,-1,r) 
	else goto node(i,z+1,1,i-1,r+1);
\end{verbatim}
\item For $j\in\{1,\dots,i-2\}$ we have:
\begin{verbatim}
node(i,z,1,j,i-1):
 if x[z]=0 then goto node(i,z+1,1,j+1,i-1) 
 else goto node(i,z+1,3,-1,j);
\end{verbatim}
\item 
\begin{verbatim}
node(i,z,1,i-1,i-1): 
if x[z]=0 then goto node(i,z+1,2,-1,i-1) 
else goto node(i,z+1,3,-1,i-1);
\end{verbatim}
\item For $q\in\{0,1\}, j\in\{1,\dots, i-1\}$ we have:
\begin{verbatim}
node(i,z,2,q,j):
 if x[z]=0 then goto node(i,z+1,2,q,j) 
 else goto node(i,z+1,2,q,j+1);
\end{verbatim}
 \item For $q\in\{0,1\}, r\in\{1,\dots, i-1\}$ we have:
\begin{verbatim}
node(i,z,3,q,r):
 if x[z]=0 then goto node(i,z+1,3,q,r+1) 
 else goto node(i,z+1,3,q,r);
\end{verbatim}
\item
\begin{verbatim}
node(i,z,reject):
if x[z]=0 then goto node(i,z+1,reject) 
else goto node(i,z+1,reject);
node(i,z,equals):
if x[z]=0 then goto node(i,z+1,equals) 
else goto node(i,z+1,equals);
\end{verbatim}
\end{itemize}

If $x_z$ is {\em value} bit then we have following program: 
\begin{itemize}
\item
For $j,r\in\{1,\dots,i-1\}$ we have:
\begin{verbatim}
node(i,z,1,j,r):
if x[z]=0 then goto node(i,z+1,1,j,r) 
else goto node(i,z+1,1,j,r);
\end{verbatim}
\item
For $q\in\{0,1\}, j\in\{1,\dots, i-1\}$ we have:
\begin{verbatim}
node(i,z,2,q,j):
if x[z]=0 then goto node(i,z+1,2,q,j) 
else goto node(i,z+1,2,q,j);
\end{verbatim}
\item
For $j\in\{1,\dots, i-1\}$ we have:
\begin{verbatim}
node(i,z,2,-1,j):
if x[z]=0 then goto node(i,z+1,2,0,j) 
else goto node(i,z+1,2,1,j);
\end{verbatim}
\item
For $q\in\{0,1\}$ we have:
\begin{verbatim}
node(i,z,2,q,i):
if x[z]=q then goto node(equals) 
else goto node(reject);
\end{verbatim}
\item
For $q\in\{0,1\},r\in\{1,\dots, i-1\}$ we have:
\begin{verbatim}
node(i,z,3,q,j):
if x[z]=0 then goto node(i,z+1,3,q,r) 
else goto node(i,z+1,3,q,r);
\end{verbatim}
\item
For $r\in\{1,\dots, i-1\}$ we have:
\begin{verbatim}
node(i,z,3,-1,r):
if x[z]=0 then goto node(i,z+1,3,0,r) 
else goto node(i,z+1,3,1,r);
\end{verbatim}
\item
For $q\in\{0,1\}$ we have:
\begin{verbatim}
node(i,z,3,q,i):
if x[z]=q then goto node(i,z+1,equals) 
else goto node(i,z+1,reject);
\end{verbatim}
\item
\begin{verbatim}
node(i,z,reject):
if x[z]=0 then goto node(i,z+1,reject) 
else goto node(i,z+1,reject);
node(i,z,equals):
if x[z]=0 then goto node(i,z+1,equals) 
else goto node(i,z+1,equals);
\end{verbatim}
\end{itemize}

Let us describe the code of the layer $i$, level $n$ (in code we use $x[z]$ for $x_z$), for $i\in\{1,\dots,k-1\}$:
\begin{itemize}
\item For $q\in\{0,1\}$ we have:
\begin{verbatim}
node(i,n,equals):
if x[n]=0 then goto node(i+1,1,1,0,0) 
else goto node(i+1,1,1,0,0);
node(i,n,2,q,i):
if x[n]=q then goto node(i+1,1,1,0,0) 
else goto node(i+1,1,reject);
node(i,n,3,q,i):
if x[n]=q then goto node(i+1,1,1,0,0) 
else goto node(i+1,1,reject);
\end{verbatim}
\item for all other nodes we have:
\begin{verbatim}
if x[n]=0 then goto node(i+1,1,reject) 
else goto nodenode(i+1,1,reject) 
\end{verbatim}
\end{itemize}

Let us describe the code of the layer $k$, level $n$:
\begin{itemize}
\item For $q\in\{0,1\}$ we have:
\begin{verbatim}
node(i,n,equals):
if x[n]=0 then goto 1-sink 
else goto 1-sink;
node(i,n,2,q,i):
if x[n]=q then goto 1-sink
else goto 0-sink;
node(i,n,3,q,i):
if x[n]=q then goto 1-sink 
else goto 0-sink;
\end{verbatim}
\item for all other nodes we have:
\begin{verbatim}
if x[n]=0 then goto 0-sink 
else goto 0-sink
\end{verbatim}
\end{itemize}

Program $P$ computes $EQS_k$ by construction.

Let us compute width of $P$. We know that $i\leq k$, hence number of the first group nodes does not great $k^2$, the second and the third group does not great $3k$ and the forth group does not great $2$. Therefore width of $P$ does not great $k^2 + 6k+2$.
\Endproof 

Boolean function $EQS_k\not\in (k/r)\bf{\mbox{-}NOBDD}_{C}$ due to Lemma \ref{peq2_knobdd}. But $EQS_k\in k\bf{\mbox{-}OBDD}_{k^2 + 6k+2}$, due to Lemma \ref{peq1}. We have $(k^2 + 6k+2)\in C$ hence $EQS_k\in k\bf{\mbox{-}OBDD}_C$ and $k{\bf\mbox{-}OBDD}_{C}\subseteq k{\bf\mbox{-}NOBDD}_{C}$, consequently $EQS_k\in  k{\bf\mbox{-}NOBDD}_{C}$. These statements prove the claim of Theorem \ref{th-hi}. 
\Endproof

We have the similar theorem for deterministic case:

\begin{theorem}\label{th-hi2}
 For $k=k(n)$ and $w=w(n)$ such that $k\log w = o(n)$, $w\in C$ for set $C\in\{{\tt poly, superpoly_\alpha, subexp_\alpha}\}$, the following inclusion is true:
\[
\left\lfloor k /r \right\rfloor{\bf\mbox{-}OBDD}_{C}\subsetneq k{\bf\mbox{-}OBDD}_{C}
\]
for $\log w'=o(r) $, $r<k$  for any $w'\in C$
\end{theorem}
\Beginproof
Clearly we have that $\left\lfloor k /r\right\rfloor{\bf\mbox{-}OBDD}_{C}\subseteq  k{\bf\mbox{-}OBDD}_{C}$. Let us prove inequality of these classes. We prove it the same way we have depicted Theorem  \ref{th-hi}. $EQS_k\not\in  (k/r){\bf\mbox{-}NOBDD}_{C}$ and hence $EQS_k\not\in (k/r){\bf\mbox{-}OBDD}_{C}$, because $k$-OBDD is particular case of $k$-NOBDD. And $EQS_k\in k{\bf\mbox{-}OBDD}_{C}$, therefore $\left\lfloor k/r\right\rfloor{\bf\mbox{-}OBDD}_{C}\neq k{\bf\mbox{-}OBDD}_{C}$.
\Endproof

We have the following corollaries from above mentioned theorems:


\begin{corollary}
\label{hierarchyResult1}
For  $k = o(n/\log n)$ and $\log^2 n = o(k)$ the following statement is true:
\[ \mbox{NP-}\left(k/\log^2 n\right)\mbox{OBDD}\subsetneq \mbox{NP-}k\mbox{OBDD}. \]

For  $k = o(n/\log^{\alpha} n)$ and $\log^{\alpha+1}n = o(k)$ the following statement is true:
\[\left\lfloor k /\log^{\alpha+1}n\right\rfloor{\bf\mbox{-}{NOBDD}}_{{\tt superpoly}_\alpha}\subsetneq k{\bf\mbox{-}{NOBDD}}_{{\tt superpoly}_\alpha}.\]

For  $k = o(n^{1-\alpha})$ and $n^{\alpha}\log n = o(k)$ the following statement is true:
\[\left\lfloor k /(n^{\alpha}\log n)\right\rfloor{\bf\mbox{-}{NOBDD}}_{{\tt subexp}_\alpha}\subsetneq k{\bf\mbox{-}{NOBDD}}_{{\tt subexp}_\alpha}.\]

\end{corollary}

\begin{corollary}
\label{hierarchyResult2}
For  $k = o(n/\log n)$ and $\log^2 n = o(k)$ the following statement is true:

\[ \mbox{P-}\left(k/\log^2 n\right)\mbox{OBDD}\subsetneq \mbox{P-}k\mbox{OBDD}. \]
For  $k = o(n/\log^{\alpha} n)$ and $\log^{\alpha+1}n = o(k)$ the following statement is true:
\[\left\lfloor k /(\log^{\alpha+1}n)\right\rfloor{\bf\mbox{-}OBDD}_{{\tt superpoly}_\alpha}\subsetneq k{\bf\mbox{-}OBDD}_{{\tt superpoly}_\alpha}.\]
For  $k = o(n^{1-\alpha})$ and $n^{\alpha}\log n = o(k)$ the following statement is the:
\[\left\lfloor k /(n^{\alpha}\log n)\right\rfloor{\bf\mbox{-}OBDD}_{{\tt subexp}_\alpha}\subsetneq k{\bf\mbox{-}OBDD}_{{\tt subexp}_\alpha}.\]
\end{corollary}

\subsection{Results for Sublinear Width}
\begin{theorem}\label{th-hi-small}
 For $k=k(n), w=w(n)$ and $r=r(n)$ such that $kw(\log_2 w)=o(n)$, $k>4, w>20, r>\frac{48v\log_2 v}{w\log_2 w}$, $w,v\in C$ for set $C\in\{{\tt const, superpolylog, sublinear}\}$, the following inclusion is true:
$
\left\lfloor k/r\right\rfloor{\bf\mbox{-}OBDD}_{C}\subsetneq k{\bf\mbox{-}OBDD}_{C}.
$
\end{theorem}
\Beginproof
Clearly, we have that $\left\lfloor k/r\right\rfloor{\bf\mbox{-}OBDD}_{C}\subseteq k{\bf\mbox{-}OBDD}_{C}$. In order to prove the proper inclusion we use properties of the Boolean function $SAF_{k,w}$.

\begin{lemma}[\cite{k2015}]\label{peq1-small}
There is $2k$-OBDD $P$ of width $3w+1$ which computes $SAF_{k,w}$
\end{lemma}

Boolean function $SAF_{\lfloor k/2\rfloor,\lfloor (w-1)/3 \rfloor}\not\in (k/r){\bf\mbox{-}OBDD}_{C}$ due to Lemma \ref{peq2_knobdd-small}. But $SAF_{\lfloor k/2\rfloor,\lfloor (w-1)/3 \rfloor}\in k{\bf\mbox{-}OBDD}_{w}$, due to Lemma \ref{peq1-small}. We have $((w-1)/3)\in C$ hence $SAF_{\lfloor k/2\rfloor,\lfloor (w-1)/3 \rfloor}\in k{\bf\mbox{-}OBDD}_C$. This proves the claim of the theorem. 
\Endproof

We have the similar theorem for nondeterministic case:

\begin{theorem}\label{th-hi2-small}
 For $k=k(n), w=w(n)$ and $r=r(n)$ such that $kw^2=o(n)$, $k>4, w>20, r>\frac{48v^2}{w\log_2 w}$, $w\in C$ for set $C\in\{{\tt const, superpolylog, sublinear}\}$, the following inclusion is right:
$
\left\lfloor k /r \right\rfloor{\bf\mbox{-}NOBDD}_C\subsetneq k{\bf\mbox{-}NOBDD}_C
$
\end{theorem}
\Beginproof
Clearly, we have that $\left\lfloor k /r\right\rfloor{\bf\mbox{-}NOBDD}_C\subseteq k{\bf\mbox{-}NOBDD}_C$. Let us prove inequality of these classes. We prove it the same way we have depicted Theorem  \ref{th-hi-small}. $SAF_{\lfloor k/2\rfloor,\lfloor (w-1)/3 \rfloor}\not\in (k/r){\bf\mbox{-}NOBDD}_C$  due to Lemma \ref{peq2_knobdd-small}. And $SAF_{\lfloor k/2\rfloor,\lfloor (w-1)/3 \rfloor}\in k{\bf\mbox{-}NOBDD}_C$ according to Lemma \ref{peq1-small}, therefore $\left\lfloor k /r \right\rfloor{\bf\mbox{-}NOBDD}_C\neq k{\bf\mbox{-}NOBDD}_C$.
\Endproof

And similar one for probabilistic case:

\begin{theorem}\label{th-hi3-small}
 For $k=k(n), w=w(n)$ and $r=r(n)$ such that $kw^2\log(k(\log w + \log k)) =o(n)$, $k>4, w>20, w^2\log (k(\log k + \log w))=o(r)$, $w\in C$ for set $C\in\{{\tt const, superpolylog, sublinear}\}$, the following inclusion is right:
$
\left\lfloor k /r \right\rfloor{\bf\mbox{-}POBDD}_C\subsetneq k{\bf\mbox{-}POBDD}_{C}
$
\end{theorem}
\Beginproof
Clearly, we have that $\left\lfloor k /r\right\rfloor{\bf\mbox{-}POBDD}_C\subseteq k{\bf\mbox{-}POBDD}_C$. Let us prove inequality of these classes. We prove it the same way we have depicted Theorem  \ref{th-hi-small}. $SAF_{\lfloor k/2\rfloor,\lfloor (w-1)/3 \rfloor}\not\in (k/r){\bf\mbox{-}POBDD}_C$ due to Lemma \ref{peq2_kpobdd-small}. And $SAF_{\lfloor k/2\rfloor,\lfloor (w-1)/3 \rfloor}\in k{\bf\mbox{-}POBDD}_C$ according to Lemma \ref{peq1-small}, therefore $\left\lfloor k /r \right\rfloor{\bf\mbox{-}POBDD}_C\neq k{\bf\mbox{-}POBDD}_C$.
\Endproof

We have the following corollaries from above mentioned theorems:


\begin{corollary}
\label{hierarchyResult1-small}
For  $k =o(n/\log n)$ and $\log n = o(k)$ the following statement is true:
\[ \left\lfloor k/(\log_2\log_2 n)\right\rfloor {\bf\mbox{-}OBDD}_{{\tt const}}\subsetneq k{\bf\mbox{-}OBDD}_{{\tt const}}. \]

For  $\varepsilon,\varepsilon_1>0, k =o(n^{1-\varepsilon}), n^{\varepsilon_1}<k$ the following statement is true:
\[\left\lfloor k/n^{\varepsilon_1}\right\rfloor{\bf\mbox{-}OBDD}_{{\tt polylog}}\subsetneq k{\bf\mbox{-}OBDD}_{{\tt polylog}}.\]

For  $0<\alpha<0.5-\varepsilon, \varepsilon>0, k>n^{\alpha}(\log_2 n)^2, k =o(n^{1-\alpha}/\log_2 n)$ the following statement is true:
\[\left\lfloor k/(n^{\alpha}(\log_2 n)^2)\right\rfloor{\bf\mbox{-}OBDD}_{{\tt sublinear}_{\alpha}}\subsetneq k{\bf\mbox{-}OBDD}_{{\tt sublinear}_{\alpha}}.\]

\end{corollary}

\begin{corollary}
\label{hierarchyResult2-small}
For  $k =o(n/\log n)$ and $\log n = o(k)$ the following statement is true:
\[ \left\lfloor k/(\log_2\log_2 n)\right\rfloor {\bf\mbox{-}NOBDD}_{{\tt const}}\subsetneq k{\bf\mbox{-}NOBDD}_{{\tt const}}. \]

For  $\varepsilon,\varepsilon_1>0, k =o(n^{1-\varepsilon}), n^{\varepsilon_1}<k$ the following statement is true:
\[\left\lfloor k/n^{\varepsilon_1}\right\rfloor{\bf\mbox{-}NOBDD}_{{\tt polylog}}\subsetneq k{\bf\mbox{-}NOBDD}_{{\tt polylog}}.\]

For  $0<\alpha<0.25-\varepsilon,\varepsilon>0, k>n^{2\alpha}(\log_2 n)^2, k =o(n^{1-2\alpha}/\log_2 n)$  following statement is right:
\[\left\lfloor k/(n^{2\alpha}(\log_2 n)^2)\right\rfloor{\bf\mbox{-}NOBDD}_{{\tt sublinear}_{\alpha}}\subsetneq k{\bf\mbox{-}NOBDD}_{{\tt sublinear}_{\alpha}}.\]
\end{corollary}

\begin{corollary}
\label{hierarchyResult3-small}
For  $k =o(n/\log n)$ and $\log_2n \cdot \log_2\log_2 n= o(k)$ the following statement is true:
\[ \left\lfloor k/(\log_2n \cdot \log_2\log_2 n)\right\rfloor {\bf\mbox{-}POBDD}_{{\tt const}}\subsetneq k{\bf\mbox{-}POBDD}_{{\tt const}}. \]

For  $\varepsilon, \varepsilon_1>0, k =o(n^{1-\varepsilon}), n^{\varepsilon_1}<k$ the following statement is true:
\[\left\lfloor k/n^{\varepsilon_1}\right\rfloor{\bf\mbox{-}POBDD}_{{\tt polylog}}\subsetneq k{\bf\mbox{-}POBDD}_{{\tt polylog}}.\]

For  $0<\alpha<0.25-\varepsilon,\varepsilon>0, k>n^{2\alpha}(\log_2 n)^2, k =o(n^{1-2\alpha}/(\log_2 n)^2)$  following statement is right:
\[\left\lfloor k/(n^{2\alpha}(\log_2 n)^2)\right\rfloor{\bf\mbox{-}POBDD}_{{\tt sublinear}_{\alpha}}\subsetneq k{\bf\mbox{-}POBDD}_{{\tt sublinear}_{\alpha}}.\]
\end{corollary}

\section*{Acknowledgements}
Partially supported by Russian Foundation for Basic Research,
Grants 14-07-00557, 14-07-00878 and 15-37-21160. The work is performed according to the Russian Government Program of Competitive Growth of Kazan Federal University

\end{document}